\documentclass[12pt, reqno]{amsart}
\usepackage{amssymb}
\usepackage{mathrsfs}
\usepackage{mathpazo}
\usepackage[colorlinks=true, breaklinks=true]{hyperref}
\usepackage{bm}
\usepackage[latin1]{inputenc}
\usepackage{graphicx}
\newtheorem{theorem}{Theorem}

\newtheorem{corollary}[theorem]{Corollary}

\newtheorem{definition}[theorem]{Definition}

\newtheorem{lemma}[theorem]{Lemma}

\newtheorem{proposition}[theorem]{Proposition}
\newtheorem{remark}[theorem]{Remark}

\newcommand{\bea}{\begin{eqnarray}}
\newcommand{\eq}{\end{eqnarray}}
\newcommand{\eea}{\end{eqnarray}}
\newcommand{\bqn}{\begin{eqnarray*}}
\newcommand{\beaa}{\begin{eqnarray*}}
\newcommand{\eqn}{\end{eqnarray*}}
\newcommand{\eeaa}{\end{eqnarray*}}
\newcommand{\bpr}{\begin{proposition}}
\newcommand{\epr}{\end{proposition}}

\newcommand{\cal}{\mathcal}

\numberwithin{equation}{section}
\numberwithin{theorem}{section}
\begin{document}
\title{Gaussian stochastic volatility models: Scaling regimes, large deviations, and moment explosions}
\author{Archil Gulisashvili}\let\thefootnote\relax\footnotetext{Department of Mathematics, Ohio University, Athens OH 45701; e-mail: gulisash@ohio.edu}
\date{}

\begin{abstract}
In this paper, we establish sample path large and moderate deviation principles for log-price processes in Gaussian stochastic volatility models, and study the asymptotic behavior of exit probabilities, call pricing functions, and the implied volatility. In addition, 
we prove that if the volatility function in an uncorrelated Gaussian model grows faster than linearly, then, for the asset price process, all the moments of order greater than one are infinite. Similar moment explosion results are obtained for correlated models. 
\end{abstract}

\maketitle

\vspace{0.2in%
}

 \vspace{%
0.2in}

\section{Introduction}\label{S:in}
This paper deals with Gaussian stochastic volatility models. In such a model, the volatility process is a positive function $\sigma$ of a Gaussian process $\widehat{B}$. The main results obtained in the paper are the following (see the end of the introduction for a more detailed overview):
\begin{itemize}
\item A sample path large deviation principle for the log-price process in a Volterra type Gaussian stochastic volatility model 
(see Theorem \ref{T:2}) with application to the exit time probability function asymptotics (see Theorem \ref{T:33}). 
\item A sample path moderate deviation principle for the log-price process in a Gaussian stochastic volatility model (see Theorem \ref{T:tet}). 
\item The results in Section \ref{S:revisit} concerning moment explosions for asset price processes in Gaussian stochastic volatility models, especially Theorem \ref{T:confitur}.
\end{itemize}
In the present paper, we also suggest a unified approach to various scaling regimes associated with Gaussian stochastic volatility models. 
More precisely, large deviation, moderate deviation, and central limit scalings are considered. Sample path large and moderate deviation 
principles are established in this paper under very mild restrictions on the volatility function and the volatility process. We also find leading terms in asymptotic expansions of call pricing functions and the implied volatility in mixed scaling regimes. To find more terms in such expansions, additional smoothness restrictions have to be imposed on the volatility function $\sigma$ 
(see, e.g., \cite{BFGHS}). Higher order expansions of call pricing functions and the implied volatility are not discussed in the present paper.
We refer the interested reader to an important paper \cite{FGPi}, where such expansions are studied.

The asset price process $S$ in a Gaussian stochastic volatility model satisfies the following stochastic differential equation:
\begin{equation}
dS_t=S_t\sigma(\widehat{B}_t)dZ_t,\quad S_0=s_0> 0,\quad 0\le t\le T,
\label{E:DD}
\end{equation}
where $s_0$ is the initial price, and $T> 0$ is the time horizon. The process $Z$ in (\ref{E:DD}) is standard Brownian motion. The equation in (\ref{E:DD}) is considered on a filtered probability space $(\Omega,\mathcal{F},\{\mathcal{F}_t\}_{0\le t\le T},\mathbb{P})$, where $\{\mathcal{F}_t\}_{0\le t\le T}$ is the augmentation of the filtration generated by the process $Z$ (see \cite{KaS}, Definition 7.2). The filtration $\{\mathcal{F}_t\}$ is right-continuous (\cite{KaS}, Corollary 7.8). It is assumed in (\ref{E:DD}) that $\sigma$ is a nonnegative continuous function on $\mathbb{R}$, and $\widehat{B}$ is a nondegenerate continuous Gaussian process adapted to the
filtration $\{\mathcal{F}_t\}_{0\le t\le T}$. In Section \ref{S:ldr} devoted to large deviation principles, the process $\widehat{B}$ is a continuous Gaussian process possessing a Volterra type representation with respect to the process $B$. It follows from (\ref{E:DD}) that the evolution of volatility in a Gaussian stochastic volatility model is described by the stochastic process $\sigma(\widehat{B})$. 
The function $\sigma$ and the process $\widehat{B}$ will be called the volatility function and the volatility process, respectively. 

We will often need to take into account the correlation structure between the asset price and the volatility. 
It will be assumed in such a case that standard Brownian motion $Z$, appearing in (\ref{E:DD}), has the following form: 
$Z_t=\bar{\rho}W_t+\rho B_t$, where $W$ and $B$ are independent standard Brownian motions, $\rho\in[-1,1]$ is the correlation coefficient, and  $\bar{\rho}=\sqrt{1-\rho^2}$. Then, the model for the asset price takes the following form:
\begin{equation}
dS_t=S_t\sigma(\widehat{B}_t)(\bar{\rho}dW_t+\rho dB_t),\quad S_0=s_0> 0,\quad 0\le t\le T.
\label{E:mood}
\end{equation}
In the special case, where the correlation coefficient $\rho$ equals zero, the model in (\ref{E:mood}) is called uncorrelated.
The asset price process in such a model satisfies the stochastic differential equation
$$
dS_t=S_t\sigma(\widehat{B}_t)dW_t,\,\,S_0=s_0,\,\,0\le t\le T.
$$

Let us denote by $\{\widetilde{\mathcal{F}}_t\}_{0\le t\le T}$ the augmentation of the filtration generated by the process $B$. If the volatility process $\widehat{B}$ is a Volterra type continuous Gaussian process (see 
Definitions \ref{D:Volt1} and \ref{D:Volt2} in Section \ref{S:ldr}), then it is adapted to the filtration 
$\{\widetilde{\mathcal{F}}_t\}_{0\le t\le T}$, and the model in (\ref{E:mood}) looks like a classical 
correlated stochastic volatility model. We call such a model a Volterra type Gaussian stochastic volatility model.
Note that Definition \ref{D:Volt2} of a Volterra type process with H\"{o}lder kernel includes an $r$-H\"{o}lder-type condition in $L^{2}$ for the kernel of the volatility process.

The unique solution to the equation in (\ref{E:DD}) is the Dol\'{e}ans-Dade exponential
$$
S_t=s_0\exp\left\{-\frac{1}{2}\int_0^t\sigma^2(\widehat{B}_s)ds+\int_0^t\sigma(\widehat{B}_s)dZ_s\right\},\quad 0\le t\le T.
$$
Therefore, the log-price process $X_t=\log S_t$ satisfies
\begin{equation}
X_t=x_0-\frac{1}{2}\int_0^t\sigma^2(\widehat{B}_s)ds+\int_0^t\sigma(\widehat{B}_s)dZ_s,
\label{E:poz}
\end{equation}
where $x_0=\log s_0$. 

Suppose $H> 0$, $\beta\in[0,H]$, and let $\varepsilon\in(0,1]$ be a small-noise parameter. 
We will work with the following scaled version of the model in (\ref{E:DD}):
$$
dS^{\varepsilon,\beta,H}_t=\varepsilon^{H-\beta}S^{\varepsilon,\beta,H}_t\sigma\left(\varepsilon^{H}\widehat{B}_t\right)dZ_t,
$$
where $0\le t\le T$. For the sake of simplicity, we often assume that the initial condition $s_0$ for the asset price satisfies $s_0=1$. The asset price process in the scaled model is given by
\begin{align}
&S_t^{\varepsilon,\beta,H}=\exp\left\{-\frac{1}{2}\varepsilon^{2H-2\beta}\int_0^t\sigma(\varepsilon^{H}\widehat{B}_s)^2ds
+\varepsilon^{H-\beta}
\int_0^t\sigma(\varepsilon^{H}\widehat{B}_s)dZ_s\right\},\quad 0\le t\le T,
\label{E:DDES}
\end{align}
while the log-price process is as follows:
\begin{equation}
X^{\varepsilon,\beta,H}_t=-\frac{1}{2}\varepsilon^{2H-2\beta}\int_0^t\sigma(\varepsilon^{H}\widehat{B}_s)^2ds+\varepsilon^{H-\beta}
\int_0^t\sigma(\varepsilon^{H}\widehat{B}_s)dZ_s,\quad 0\le t\le T.
\label{E:eq1}
\end{equation}
It is easy to understand how the results obtained in the present paper transform if $s_0\neq 1$. One can simply replace the process 
$X^{\varepsilon,\beta,H}$ by the process $X^{\varepsilon,\beta,H}-x_0$.

We will next provide a brief overview of the results obtained in the paper. Sections \ref{S:ldr} and \ref{S:mdr} are devoted to sample path large and moderate deviation principles for log-price processes. The theory of sample path LDPs for solutions of stochastic differential equations goes back to a celebrated work of Freidlin and Wentzell (see \cite{FW}; for more information consult \cite{DZ,DS,V}). We also refer the reader to \cite{BC,CMD,Ph,R} for applications of sample path large deviation principles in financial mathematics.

In the case where $\beta=0$, the model is in the large deviation scaling regime. In Section \ref{S:ldr}, we prove a sample path large deviation principle (LDP) 
for the log-price process $\varepsilon\mapsto X^{\varepsilon,0,H}$ (see Theorem \ref{T:2}). A similar sample path LDP was obtained in a recent pre-print \cite{CP} of Cellupica and Pacchiarotti under more restrictive assumptions on the volatility function $\sigma$. It is common to apply small-noise sample path large deviation principles in the study of the asymptotic behavior of the exit probabilities. Such results go back to the work of Freidlin and Wentzell (see \cite{FV11,FV12,FW}). Different proofs of these results, using stochastic control theory, were given by Fleming in \cite{Fl}. In Section 
\ref{S:ldr}, we characterize the leading term in the asymptotic expansion of the exit time probability function, using the large deviation principle obtained in Theorem \ref{T:2} (see Theorem \ref{T:33}). A similar result was obtained in \cite{CP} under more restrictions on the volatility function $\sigma$. Note that a large deviation principle for the process $\varepsilon\mapsto X^{\varepsilon,0,H}_T$ with state space $\mathbb{R}$ was earlier established in Forde and Zhang \cite{FZ} in the case, where the function 
$\sigma$ satisfies the global H\"{o}lder condition, while the process $\widehat{B}$ is fractional Brownian motion. In \cite{G}, we proved the 
Forde-Zhang LDP under very mild restrictions on $\sigma$ and $\widehat{B}$. We formulate the latter result 
in Section \ref{S:ldr}. 

If $0<\beta< H$, then the model is in the moderate deviation scaling regime (see, e.g., \cite{BFGHS,EL,FGP}, and the references therein for more information on moderate deviations). In Section \ref{S:mdr}, we prove a sample path moderate deviation principle (MDP) for the process $\varepsilon\mapsto X^{\varepsilon,\beta,H}$ (see Theorem \ref{T:tet}), and derive a corresponding MDP for the process 
$\varepsilon\mapsto X^{\varepsilon,\beta,H}_T$ (see Corollary \ref{C:corio}). As it often happens in the theory of moderate deviations, the rate function in Corollary \ref{C:corio} is quadratic. At the end of Section \ref{S:mdr}, we explain how to pass from small-noise large and moderate deviation principles to small-time ones under the condition that the volatility process is self-similar. 

The case where $\beta=H$ corresponds to the central limit (CL) scaling regime. In Section \ref{S:clr}, we characterize the limiting behavior on the path space of the distribution function of the process $\varepsilon\mapsto X^{\varepsilon,H,H}$ (see Theorem \ref{T:theor}), and also that of the process $\varepsilon\mapsto X^{\varepsilon,H,H}_T$ in the space $\mathbb{R}$ (see Theorem \ref{T:clr}). The results in the CL regime can be considered as degenerate MDPs with the rate function equal to a constant (see Remark \ref{R:uz1} in Section \ref{S:clr}). An example of a CL scaling can be found in \cite{GaS1}. The volatility of an asset in \cite{GaS1} is modeled by the process 
$\delta\mapsto\sigma(\delta\widetilde{U})$, where $\sigma$ is a smooth function, while $\widetilde{U}$ is the stationary fractional Ornstein-Uhlenbeck process (our notation is different from that used in Section 3 of \cite{GaS1}). The CL scaling in \cite{GaS1} corresponds to the following values of the parameters: $H=1$ and $\beta=1$ (our notation).

It follows from what was said above that the class of small-noise parametrizations of the log-price process in a 
Gaussian stochastic volatility model (see formula (\ref{E:DDES})) can be split into three disjoint subclasses, which correspond to large deviation, moderate deviation, and central limit scaling regimes. An interesting discussion of certain differences between those regimes can be 
found in \cite{EL}. Gaussian stochastic volatility models and their scaled versions were studied in 
\cite{BFGHS,FZ,FGPi,FGP,G,GaS1,GS,GVZ1,GVZ2}. 
In the last years, fractional Gaussian stochastic volatility models have become increasingly popular. The volatility process in such 
a model is a fractional Gaussian process, for instance, fractional Brownian motion, the Riemann-Liouville fractional Brownian motion, or the fractional Ornstein-Uhlenbeck process (see the next section for the definitions of these processes). 
We refer the reader to short surveys of Gaussian fractional stochastic volatility models in \cite{G,GS,GaS1} for more information. 
A unified approach to LDP and MDP regimes in fractional stochastic volatility models is suggested in \cite{FGPi}. 

In Section \ref{S:varre}, we find leading terms in asymptotic expansions of call pricing functions in mixed scaling regimes. The methods allowing to derive call price estimates from the validity of a large deviation principle for the log-price are well-known. Such derivations often exploit the linear growth condition for the volatility function and the finiteness of either the moments of the asset price process, or the exponential moments of the integrated variance (see the discussion in Section \ref{S:varre}). However, in Section \ref{S:revisit}, we show that if the volatility function grows slightly faster than the first power, then all the nontrivial exponential moments of the integrated variance are infinite (see Theorem \ref{T:confit}). Moreover, it is established that in an uncorrelated Gaussian stochastic 
volatility model with the volatility function growing faster than linearly, all the moments of order greater than one of the asset price process are infinite (see part (i) of Theorem \ref{T:confitur}). The previous assertion was first included in the arXiv:1808.00421v4 (September 22, 2018) version of the present paper. The same result in a special case, where the volatility process is the Riemann-Liouville fractional Brownian motion with the Hurst index $H>\frac{1}{2}$, while the volatility function satisfies an additional condition 
(condition (G) formulated before Theorem \ref{T:confitur}), was obtained independently, but a little later, by Gassiat (see Theorem 2 in the arXiv:1811.10935v1 (November 27, 2018) version of \cite{Gassiat}).
We also show that in a correlated Volterra type Gaussian 
model ($\rho\neq 0$), all the moments of the order $\gamma>\frac{1}{1-\rho^2}$ explode (see part (ii) of Theorem \ref{T:confitur}). Such a result was first established by Jourdain for the Scott model (see \cite{J}). In the Scott model, the volatility process is the Ornstein-Uhlenbeck process, while the volatility function is $\sigma(x)=e^x$. Gassiat obtained a similar result for a model with $\rho< 0$, the volatility function satisfying condition (G), and the Riemann-Liouville fractional Brownian motion with $H>\frac{1}{2}$ as the volatility process (see Theorem 2 in \cite{Gassiat}). We do not assume in part (ii) of Theorem \ref{T:confitur} that the model is negatively correlated and the volatility function satisfies condition (G). Moreover, the volatility process in Theorem \ref{T:confitur} may be any Volterra type continuous Gaussian process. In the present paper, we also obtain partial results concerning the explosion of the moment of order $\gamma=\frac{1}{1-\rho^2}$ in a correlated Gaussian stochastic volatility model (see Theorem \ref{T:finaluli}). More information can be found in Remarks \ref{R:ooo} and \ref{R:rrr} below. We would also like to bring 
the attention of the reader to the paper \cite{Pes}, where the author explains how the exponential integrability 
of the maximal function of a continuous local martingale depends on the growth of the moments of its quadratic variation. 

The last section (Section \ref{S:varren}) of the present paper is devoted to the study of small-noise asymptotic behavior of the implied volatility in Gaussian stochastic volatility models under various scaling regimes. 
\section{Large deviations: $\beta=0$}\label{S:ldr}
We have already mentioned in the introduction that in \cite{FZ}, Forde and Zhang obtained a large deviation principle for the log-price process in a fractional Gaussian stochastic volatility model, under the assumption that the volatility function satisfies a global H\"{o}lder condition, while the volatility process is fractional Brownian motion. This result was generalized in \cite{G}, where an additional scaling was introduced, and the LDP was established under milder conditions than those in \cite{FZ}. It was assumed in \cite{G} that the volatility function satisfies a very mild regularity condition, while the volatility process is a Volterra type continuous Gaussian process. 

Our next goal is to introduce Volterra type processes. We will also formulate the large deviation principle obtained in \cite{G}, and establish a sample path large deviation principle under the same restrictions as in \cite{G}. 

Suppose the model in (\ref{E:mood}) is fixed, and let $K$ be a square integrable kernel on $[0,T]^2$ such that
$
\sup_{t\in[0,T]}\int_0^T|K(t,s)|^2ds<\infty.
$
Let ${\cal K}:L^2[0,T]\mapsto L^2[0,T]$ be the linear operator defined by
$
{\cal K}h(t)=\int_0^TK(t,s)h(s)ds,
$
and let $\widehat{B}$ be a centered Gaussian process having the following representation in law:
\begin{equation}
\widehat{B}_t=\int_0^TK(t,s)dB_s,\quad 0\le t\le T,
\label{E:ooo}
\end{equation}
where $B$ is the Brownian motion appearing in (\ref{E:mood}). Such representations of Gaussian processes are called Fredholm 
representations. Actually, for every centered continuous Gaussian process there exists a Fredholm representation with Brownian motion depending on the process (see \cite{SV}, Theorem 3.1). In this paper, we assume that the process $\widehat{B}$ has a 
representation in (\ref{E:ooo}) with the process $B$ appearing in (\ref{E:mood}).

The modulus of continuity of the kernel $K$ in the space $L^2[0,T]$ is defined as follows:
$$
M(h)=\sup_{\{t_1,t_2\in[0,1]:|t_1-t_2|\le h\}}\int_0^T|K(t_1,s)-K(t_2,s)|^2ds,\quad 0\le h\le T.
$$

We will next define Volterra type processes and Volterra type processes with H\"{o}lder kernels.
\begin{definition}\label{D:Volt1}
The process in (\ref{E:ooo}) is called a Volterra type Gaussian process if the following condition holds for the kernel $K$: 
\\
(a)\,\,$K(t,s)=0$ for all $0\le t< s\le T$.
\end{definition}
\begin{definition}\label{D:Volt2}
The process in (\ref{E:ooo}) will be called a Volterra type Gaussian process with a H\"{o}lder kernel, if condition (a) is satisfied, and the following additional condition holds: 
\\
(b)\,\,There exist constants $c> 0$ and $r> 0$ such that $M(h)\le ch^r$ 
for all $h\in[0,T]$. 
\end{definition}
\begin{remark}\label{R:mofus}
Condition (a) is a typical Volterra type condition for the kernel. The smoothness condition (b) was included in the definitions of a Volterra type Gaussian process in \cite{H,Hu}. It was 
also used in \cite{G}. 
\end{remark}

We will next introduce classical fractional processes. For $0< H< 1$, 
fractional Brownian motion $B^H_t$, $t\ge 0$, is a centered Gaussian process with the covariance function given by
$$
C_H(t,s)=\frac{1}{2}\left(t^{2H}+s^{2H}-|t-s|^{2H}\right),\quad t,s\ge 0.
$$
The process $B^H$ was first implicitly considered by Kolmogorov in \cite{Ko}, and was studied by Mandelbrot and van Ness in \cite{MvN}. The constant $H$ is called the Hurst parameter. The Riemann-Liouville fractional Brownian motion
is defined as follows:
$$
R^H_t=\frac{1}{\Gamma(H+\frac{1}{2})}\int_0^t(t-s)^{H-\frac{1}{2}}
dB_s,\quad t\ge 0,
$$
where $0< H< 1$. This stochastic process was introduced by L\'{e}vy in \cite{PL}. More information about the process $R^H$ can be found in \cite{LS,Pi}. 
The fractional Ornstein-Uhlenbeck process is defined for $0< H< 1$ and $a> 0$, by the following formula:
$$
U_t^H=\int_0^te^{-a(t-s)}dB^H_s,\quad t\ge 0
$$
(see \cite{CKM,KS}). 

Fractional Brownian motion, the Riemann-Liouville fractional Brownian motion, and fractional Ornstein-Uhlenbeck process are Volterra type Gaussian processes with H\"{o}lder kernels, for which $r=2H$ (see Lemma 2 in \cite{G}). For fractional Brownian motion, the previous statement was established in \cite{Z}. We refer the reader to \cite{D,ER,H,Hu,JLP,MN} for 
more information on Volterra type processes.
\begin{remark}\label{R:nong}
We will assume throughout the paper that the Gaussian process $\widehat{B}$ is non-degenerated. This means that the variance function $v$ of $\widehat{B}$ satisfies the condition $v(s)> 0$ for all $s\in(0,T]$.
\end{remark}
\begin{remark}\label{R:assume}
Volatility processes satisfying the conditions in Definition \ref{D:Volt2} are used in the paper only in the results associated with the large deviation regime. In all the other regimes, Definition \ref{D:Volt1} is used.
\end{remark}
\begin{definition}\label{D:moc}
Let $\omega$ be an increasing modulus of continuity on $[0,\infty)$, that is, $\omega:\mathbb{R}^{+}\mapsto\mathbb{R}^{+}$ is an increasing function such that $\omega(0)=0$ and 
$\displaystyle{\lim_{s\rightarrow 0}\omega(s)=0}$. A function $\sigma$ defined on $\mathbb{R}$ is called locally $\omega$-continuous,
if for every $\delta> 0$ there exists a number $L(\delta)> 0$ such that for all $x,y\in[-\delta,\delta]$, the following inequality holds:
$
|\sigma(x)-\sigma(y)|\le L(\delta)\omega(|x-y|).
$ 
\end{definition}

A special example of a modulus of continuity is $\omega(s)=s^{\gamma}$ with $\gamma\in(0,1)$. In this case, the condition 
in Definition \ref{D:moc} is a local 
$\gamma$-H\"{o}lder condition. If $\gamma=1$, then the condition 
in Definition \ref{D:moc} is a local Lipschitz condition.

Denote by $\mathbb{C}_0[0,T]$ the space of continuous functions on the interval $[0,T]$. For a function $f\in\mathbb{C}_0[0,T]$, 
its norm is defined by
$
||f||_{\mathbb{C}_0[0,T]}=\sup_{t\in[0,T]}|f(t)|.
$ 
In the sequel, the symbol $\mathbb{H}^1_0[0,T]$ will stand for the Cameron-Martin space, consisting of absolutely continuous functions 
$f$ on $[0,T]$ such that $f(0)=0$ and $\dot{f}\in L^2[0,T]$, where $\dot{f}$ is the derivative of $f$. For a function 
$f\in\mathbb{H}^1_0[0,T]$, its norm in $\mathbb{H}^1_0[0,T]$ is defined by 
$$
||f||_{\mathbb{H}_0^1[0,T]}=\left\{\int_0^T\dot{f}(t)^2dt\right\}^{\frac{1}{2}}.
$$
The following notation will be used below:
$$
\widehat{f}(s)=\int_0^sK(s,u)\dot{f}(u)du.
$$

We will next formulate the large deviation principle for Volterra type Gaussian stochastic volatility models established in \cite{G}. We adapt the formulation to the notation used in the present paper.
\begin{theorem}\label{T:1}
Suppose $\sigma$ is a positive function on $\mathbb{R}$ that is locally $\omega$-continuous for some modulus of continuity $\omega$.
Let $H> 0$, and let $\widehat{B}$ be a Volterra type Gaussian process with a H\"{o}lder kernel. Set
\begin{equation}
I_T(x)=\inf_{f\in\mathbb{H}_0^1[0,T]}\left[\frac{\left(x-\rho\int_0^T\sigma(\widehat{f}(s))\dot{f}(s)ds\right)^2}
{2(1-\rho^2)\int_0^T\sigma(\widehat{f}(s))^2ds}+\frac{1}{2}\int_0^T\dot{f}(s)^2ds\right].
\label{E:vunz}
\end{equation}
Then the function $I_T$ is a good rate function. Moreover, a small-noise large deviation principle with speed $\varepsilon^{-2H}$ and rate function $I_T$ given by (\ref{E:vunz}) holds for the process
$\varepsilon\mapsto X_T^{\varepsilon,0,H}$, where $X_T^{\varepsilon,0,H}$ is defined by (\ref{E:eq1}). More precisely,
for every Borel measurable subset $A$ of $\mathbb{R}$, the following estimates hold:
\begin{align*}
&-\inf_{x\in A^{\circ}}I_T(x)\le\liminf_{\varepsilon\downarrow 0}\varepsilon^{2H}\log\mathbb{P}\left(
X_T^{\varepsilon,0,H}\in A\right) 
 \\
&\le\limsup_{\varepsilon\downarrow 0}\varepsilon^{2H}\log\mathbb{P}\left(X_T^{\varepsilon,0,H}\in A\right)
\le-\inf_{x\in\bar{A}}I_T(x).
\end{align*}
The symbols $A^{\circ}$ and $\bar{A}$ in the previous estimates stand for the interior and the closure of the set $A$, respectively.
\end{theorem}
\begin{remark}\label{R:lsc}
Recall that a rate function on a topological space ${\cal X}$ is a lower semi-continuous mapping $I:{\cal X}\mapsto[0,\infty]$, that is,
for all $y\in[0,\infty)$, the level set $L_y=\{x\in{\cal X}:I(x)\le y\}$ is a closed subset of ${\cal X}$. A rate function $I$ is called a good rate function if for every $y\in[0,\infty)$, the set $L_y$ is a compact subset of ${\cal X}$.
\end{remark}

We refer the reader to \cite{BFGHS,FZ,G} for more information on large deviation principles for Volterra type Gaussian stochastic volatility models.

Let us define a measurable functional $\Phi$ 
from the space $M=\mathbb{C}_0[0,T]^3$ into the space $\mathbb{C}_0[0,T]$ as follows: For 
$l\in\mathbb{H}^1_0[0,T]$ and $(f,g)\in\mathbb{C}_0[0,T]^2$ such that $f\in\mathbb{H}^1_0[0,T]$ and $g=\widehat{f}$, 
$$
\Phi(l,f,g)(t)=\bar{\rho}\int_0^t\sigma(\widehat{f}(s))\dot{l}(s)ds
+\rho\int_0^t\sigma(\widehat{f}(s))\dot{f}(s)ds,\quad 0\le t\le T.
$$
In addition, for all the remaining triples $(l,f,g)$, we set $\Phi(l,f,g)(t)=0$ for all $t\in[0,T]$. 

The next statement is a sample path large deviation principle for the process $\varepsilon\mapsto X^{\varepsilon,0,H}$ 
with state space $\mathbb{C}_0[0,T]$.
\begin{theorem}\label{T:2}
Suppose the conditions in Theorem \ref{T:1} hold. 
Then the process 
$\varepsilon\mapsto X^{\varepsilon,0,H}$ defined by (\ref{E:eq1}) satisfies the small-noise large deviation principle 
with speed $\varepsilon^{-2H}$ 
and good rate function $Q_T$ given by $Q_T(g)=\infty$, for all $g\in\mathbb{C}_0[0,T]\backslash\mathbb{H}_0^1[0,T]$, and 
\begin{align}
&Q_T(g)=\inf_{f\in\mathbb{H}_0^1[0,T]}\left[\frac{1}{2}\int_0^T\left[\frac{\dot{g}(s)-\rho\sigma(\widehat{f}(s))\dot{f}(s)}
{\bar{\rho}\sigma(\widehat{f}(s))}\right]^2ds
+\frac{1}{2}\int_0^T\dot{f}(s)^2ds\right],
\label{E:vunzil}
\end{align}
for all $g\in\mathbb{H}_0^1[0,T]$. 
The validity of the large deviation principle means that
for every Borel measurable subset ${\cal A}$ of $\mathbb{C}_0[0,T]$, the following estimates hold:
\begin{align*}
&-\inf_{g\in{\cal A}^{\circ}}Q_T(g)\le\liminf_{\varepsilon\downarrow 0}\varepsilon^{2H}\log\mathbb{P}\left(
X^{\varepsilon,0,H}\in{\cal A}\right) 
 \\
&\le\limsup_{\varepsilon\downarrow 0}\varepsilon^{2H}\log\mathbb{P}\left(X^{\varepsilon,0,H}\in{\cal A}\right)
\le-\inf_{g\in\bar{{\cal A}}}Q_T(g).
\end{align*}
\end{theorem}
\begin{remark}\label{R:cp}
A similar LDP was recently obtained in \cite{CP} under more restrictive assumptions. For instance, it is supposed in \cite{CP}
that the volatility function $\sigma$ is $\alpha$-H\"{o}lder continuous and bounded from above on $\mathbb{R}$, while we assume in Theorem
\ref{T:2} that $\sigma$ is locally $\omega$-continuous for some modulus of continuity $\omega$.
\end{remark}
\begin{remark}\label{R:san}
To make a sanity check, let us consider a special case of Theorem \ref{T:2} where $\rho=0$ and $\sigma(u)=1$ for all $u\in[0,T]$.
Then
$$
Q_T(g)=\inf_{f\in\mathbb{H}_0^1[0,T]}\left[\frac{1}{2}\int_0^T\dot{g}(s)^2ds+\frac{1}{2}\int_0^T\dot{f}(s)^2ds\right]
=\frac{1}{2}\int_0^T\dot{g}(s)^2ds,
$$
and we recover Schilder's theorem.
\end{remark}
\begin{remark}\label{R:remis}
Recall that a set ${\cal A}\subset\mathbb{C}_0[0,T]$ is called a set of continuity for the rate function $Q_T$ if
\begin{equation}
\inf_{g\in{\cal A}^{\circ}}Q_T(g)=\inf_{g\in\bar{{\cal A}}}Q_T(g).
\label{E:err}
\end{equation}
For such a set, Theorem \ref{T:2} implies that
\begin{equation}
\lim_{\varepsilon\downarrow 0}\varepsilon^{2H}
\log\mathbb{P}\left(X^{\varepsilon,0,H}\in{\cal A}\right)=-\inf_{g\in{\cal A}}Q_T(g).
\label{E:errs}
\end{equation}
\end{remark}

\it Proof of Theorem \ref{T:2}. \rm For every $g\in\mathbb{C}_0[0,T]$, set
\begin{align}
&Q_T(g) \nonumber \\
&=\inf_{l,f\in\mathbb{H}_0^1[0,T]}\left[\frac{1}{2}\left(\int_0^T\dot{l}(s)^2ds
+\int_0^T\dot{f}(s)^2ds\right):\Phi(l,f,\widehat{f})(t)=g(t),\,
t\in[0,T]\right],
\label{E:vunzi}
\end{align}
if $g$ is such that the set on the right-hand side of (\ref{E:vunzi}) is not empty, and $Q_T(g)=\infty$, otherwise.
For the sake of simplicity, we assume that $T=1$ and $s_0=1$. 

It was shown in the proof in Section 6 of \cite{G} that the process 
$\varepsilon\mapsto\varepsilon^H(W_1,B,\widehat{B})$ with state space $\mathbb{R}\times\mathbb{C}_0[0,1]^2$ satisfies
the large deviation principle with speed $\varepsilon^{-2H}$ and good rate function given by 
$$
\widetilde{I}(y,f,g)=\frac{1}{2}y^2+I(f,g),\quad y\in\mathbb{R},\quad (f,g)\in\mathbb{C}_0[0,1]^2.
$$
In the previous definition, the function $I$ is defined as follows: If $f\in\mathbb{H}_0^1[0,1]$ and $g=\widehat{f}$, then
$I(f,g)=\frac{1}{2}\int_0^1\dot{f}(s)^2ds$,  and in all the remaining cases, $I(f,g)=\infty$. Similarly, we can prove that 
the process 
$\varepsilon\mapsto\varepsilon^H(W,B,\widehat{B})$ with state space $\mathbb{C}_0[0,1]^3$ satisfies
the large deviation principle with speed $\varepsilon^{-2H}$ and good rate function given by 
$$
{\cal I}(v,f,g)=\frac{1}{2}\int_0^1\dot{v}(s)^2ds+I(f,g),\quad y\in\mathbb{R},\quad (v,f,g)\in\mathbb{C}_0[0,1]^3.
$$
Here we take into account Schilder's theorem and the fact that Brownian motions $W$ and $B$ are independent.

Using the same ideas as in Section 5 of \cite{G}, we can show that if we remove the drift term, then the LDP in Theorem 
\ref{T:2} is not affected. More precisely, this means that it suffices to prove the LDP in Theorem \ref{T:2} for the process 
$\varepsilon\mapsto\widehat{X}^{\varepsilon,0,H}$, where
\begin{equation}
\widehat{X}^{\varepsilon,0,H}_t=\varepsilon^{H}
\int_0^t\sigma(\varepsilon^{H}\widehat{B}_s)(\bar{\rho}dW_s+\rho dB_s),\quad 0\le t\le 1.
\label{E:intl}
\end{equation}

Our next goal is to show how to apply the extended contraction principle in our environment 
(see Theorem 4.2.23 in \cite{DZ} for more details concerning the extended contraction principle). 
Let us define a sequence of functionals $\Phi_m:M\mapsto\mathbb{C}_0[0,1]$, $m\ge 1$ as follows: For 
$(r,h,l)\in\mathbb{C}_0[0,1]^3$, and $t\in[0,1]$, 
\begin{align*}
&\Phi_m(r,h,l)(t) \\
&=\bar{\rho}\sum_{k=0}^{[mt-1]}
\sigma\left(l\left(\frac{k}{m}\right)\right)\left[r\left(\frac{k+1}{m}\right)-r\left(\frac{k}{m}\right)\right]
+\sigma\left(l\left(\frac{k+1}{m}\right)\right)\left[r\left(t\right)-r\left(\frac{[mt]}{m}\right)\right] \\
&\quad+\rho\sum_{k=0}^{[mt-1]}
\sigma\left(l\left(\frac{k}{m}\right)\right)\left[h\left(\frac{k+1}{m}\right)-h\left(\frac{k}{m}\right)\right]
+\sigma\left(l\left(\frac{k+1}{m}\right)\right)\left[h\left(t\right)-h\left(\frac{[mt]}{m}\right)\right].
\end{align*}
It is not hard to see that for every $m\ge 1$, the mapping $\Phi_m$ is continuous.

We will next establish that formula (4.2.24) in \cite{DZ} holds in our setting. This formula is used in the 
formulation of the extended contraction principle (see \cite{DZ}, Theorem 4.2.23).
\begin{lemma}\label{L:ecp}
For every $\zeta> 0$ and $y> 0$,
$$
\limsup_{m\rightarrow\infty}\sup_{\{(r,f)\in\mathbb{H}_0^1[0,1]^2:\frac{1}{2}\int_0^1\dot{r}(s)^2ds
+\frac{1}{2}\int_0^1\dot{f}(s)^2ds\le\zeta\}}
||\Phi(r,f,\widehat{f})-\Phi_m(r,f,\hat{f})||_{\mathbb{C}_0[0,1]^2}=0.
$$
\end{lemma}

\it Proof. \rm The proof of Lemma \ref{L:ecp} is similar to that of Lemma 21 in \cite{G}. It is not hard to see that
for all $(r,f)\in\mathbb{H}_0^1[0,1]^2$ and $m\ge 1$,
$$
\Phi_m(r,f,\hat{f})=\bar{\rho}\int_0^th_m(s,f)\dot{r}(s)ds+\rho\int_0^th_m(s,f)\dot{f}(s)ds,
$$
where  
$$
h_m(s,f)=\sum_{k=0}^{m-1}\sigma\left(\widehat{f}\left(\frac{k}{m}\right)\right)\mathbb{1}_{\{\frac{k}{m}\le s\le\frac{k+1}{m}\}},\quad
0\le s\le 1.
$$
Therefore,
\begin{align}
\Phi(r,f,\widehat{f})-\Phi_m(r,f,\hat{f})&=\bar{\rho}\int_0^t[\sigma(\widehat{f}(s))-
h_m(s,f)]\dot{r}(s)ds \nonumber \\
&\quad
+\rho\int_0^t[\sigma(\widehat{f}(s))-h_m(s,f)]\dot{f}(s)ds.
\label{E:rh}
\end{align}

For every $\eta> 0$, denote
$D_{\eta}=\{w\in\mathbb{H}_0^1[0,1]:\int_0^1\dot{w}(s)^2ds\le\eta\}$. It is not hard to see that to prove Lemma \ref{L:ecp}, it suffices to show that for all $\eta> 0$,
\begin{equation}
\limsup_{m\rightarrow\infty}\left[\sup_{f\in D_{\eta},w\in D_{\eta}}\sup_{t\in[0,1]}\left|\int_0^t[\sigma(\hat{f}(s))
-h_m(s,f)]\dot{w}(s)ds\right|\right]=0,
\label{E:inter}
\end{equation}

We have
\begin{align}
&\sup_{f\in D_{\eta},w\in D_{\eta}}\sup_{t\in[0,1]}\left|\int_0^t[\sigma(\hat{f}(s))
-h_m(s,f)]\dot{w}(s)ds\right| \nonumber \\
&\le\sup_{f\in D_{\eta},w\in D_{\eta}}\int_0^1\left|\sigma(\hat{f}(s))
-h_m(s,f)\right||\dot{f}(s)|ds\le\sqrt{\eta}\sup_{f\in D_{\eta}}\sup_{s\in[0,1]}\left|\sigma(\hat{f}(s))
-h_m(s,f)\right|.
\label{E:balk}
\end{align}
It was established in the proof of Lemma 21 in \cite{G} that 
\begin{equation}
\sup_{f\in D_{\eta}}\sup_{s\in[0,1]}\left|\sigma(\hat{f}(s))
-h_m(s,f)\right|\rightarrow 0
\label{E:bru}
\end{equation}
as $m\rightarrow\infty$ (the previous statement follows from (49) in \cite{G}). Now, it is clear that (\ref{E:balk})
and (\ref{E:bru}) imply (\ref{E:inter}).

This completes the proof of Lemma \ref{L:ecp}.

It remains to prove that the sequence of processes $\varepsilon\mapsto\Phi_m\left(\varepsilon^HW,\varepsilon^HB,\varepsilon^H\widehat{B}\right)$ with state space $\mathbb{C}_0[0,1]$ is an exponentially good approximation to the process $\varepsilon\mapsto\widehat{X}^{\varepsilon,0,H}$.
The previous statement means that for every $\delta> 0$,
\begin{align}
&\lim_{m\rightarrow\infty}\,\limsup_{\varepsilon\downarrow 0}\varepsilon^{2H}\log\mathbb{P}\left(
||\widehat{X}^{\varepsilon,0,H}-\Phi_m\left(\varepsilon^HW,\varepsilon^HB,\varepsilon^H\widehat{B}\right)||
_{\mathbb{C}_0[0,1]}>\delta\right)=-\infty.
\label{E:chto}
\end{align}
Using the definitions of $\widehat{X}^{\varepsilon,0,H}$ and $\Phi_m$, we see that in order to prove the equality in (\ref{E:chto}), 
it suffices to show that for every $\tau> 0$,
\begin{equation}
\lim_{m\rightarrow\infty}\,\limsup_{\varepsilon\downarrow 0}\varepsilon^{2H}\log\mathbb{P}\left(\varepsilon^H
\sup_{t\in[0,1]}\left|\int_0^t\sigma_s^{(m)}dB_s\right|>\delta\right)=-\infty
\label{E:tru}
\end{equation}
and
\begin{equation}
\lim_{m\rightarrow\infty}\,\limsup_{\varepsilon\downarrow 0}\varepsilon^{2H}\log\mathbb{P}\left(\varepsilon^H
\sup_{t\in[0,1]}\left|\int_0^t\sigma_s^{(m)}dW_s\right|>\delta\right)=-\infty,
\label{E:trus}
\end{equation}
where 
$$
\sigma_s^{(m)}=\sigma\left(\varepsilon^H\widehat{B}_s\right)-\sigma\left(\varepsilon^H\widehat{B}_{\frac{[mt]}{m}}\right), 
\quad 0\le s\le 1,\quad m\ge 1.
$$
The formula in (\ref{E:tru}) was established in \cite{G}, formula (53). The proof of formula (\ref{E:trus}) is similar. This completes the 
proof of (\ref{E:chto}).

Finally, by taking into account (\ref{E:chto}), Lemma \ref{L:ecp}, and applying the extended contraction 
principle (Theorem 4.2.23 in \cite{DZ}), we show that the process $\varepsilon\mapsto\widehat{X}^{\varepsilon,0,H}$ satisfies the large deviation principle 
with speed $\varepsilon^{-2H}$ and good rate function $Q_1$ (see the definition in (\ref{E:vunzi})). 
For any $T> 0$, the process $\varepsilon\mapsto\widehat{X}^{\varepsilon,0,H}$ satisfies the large deviation principle 
with speed $\varepsilon^{-2H}$ and good rate function $Q_T$ defined in (\ref{E:vunzi}). The previous statement can be established using the methods employed in the reasoning before Definition 17 in \cite{G}.

We will next prove that the function $Q_T$ satisfies the formula in (\ref{E:vunzil}). It is not hard to see that 
if $g\in\mathbb{C}_0[0,T]$, $l,f\in\mathbb{H}_0^1[0,T]$, and $\Phi(l,f,\widehat{f})(t)=g(t)$ for all $t\in[0,T]$, 
then $g\in\mathbb{H}_0^1[0,1]$. Moreover, if for $g\in\mathbb{H}_0^1[0,T]$ the previous equality holds, then it is not hard to see that
$$
\dot{l}(t)=\frac{\dot{g}(s)-\rho\sigma(\widehat{f}(s)))\dot{f}(s)}
{\bar{\rho}\sigma(\widehat{f}(s))}.
$$
Now it is clear that formula (\ref{E:vunzil}) holds for the function $Q_T$ defined by (\ref{E:vunzi}).

This completes the proof of Theorem \ref{T:2}.

Our next goal in the present section is to prove that for every $g\in\mathbb{H}_0^1[0,T]$, there exists at least one minimizer $f_g$ in the minimization problem on the right-hand side of (\ref{E:vunzil}). 
\begin{lemma}\label{L:dobb}
For every function $g\in\mathbb{H}_0^1[0,T]$ there exists a function $f_g\in\mathbb{H}_0^1[0,T]$ such that
\begin{align}
&Q_T(g)=\frac{1}{2}\int_0^T\left[\frac{\dot{g}(s)-\rho\sigma(\widehat{f_g}(s))\dot{f}_g(s)}
{\bar{\rho}\sigma(\widehat{f_g}(s))}\right]^2ds
+\frac{1}{2}\int_0^T\dot{f}_g(s)^2ds.
\label{E:vunzill}
\end{align}
\end{lemma}

\it Proof. \rm It is not hard to see that it suffices to prove that for every $j\in L^2[0,T]$, the following minimization problem on the space $L^2[0,T]$ has a solution: $H(j)=\inf_{h\in L^2[0,T]}\widetilde{H}_j(h)$, where
\begin{equation}
\widetilde{H}_j(h)=\int_0^T\left[\frac{j(s)}{\bar{\rho}\sigma\left({\cal K}h(s)\right)}-\frac{\rho}{\bar{\rho}}h(s)
\right]^2ds+\int_0^Th(s)^2ds.
\label{E:eggy}
\end{equation}
In (\ref{E:eggy}), ${\cal K}h(s)=\int_0^TK(s,u)h(u)du$, and $K$ is a Volterra type kernel satisfying condition (b) in Definition \ref{D:Volt2}.
It is known that a solution to the minimization problem formulated above exists if the functional $\widetilde{H}_j:L^2[0,T]\mapsto
\mathbb{R}$ is coercive and weakly sequentially lower semi-continuous (see, e.g., \cite{St}, Ch. 1, Theorem 1.2). If the latter property holds for the functional $F$, then, for the sake of shortness, we will write $F\in WLS$. The coercivity of the functional in (\ref{E:eggy}) is clear. It is also a known fact that the 
functional $h\mapsto\int_0^Th(s)^2ds$ belongs to the class $WLS$. Since the sum of two  functionals from $WLS$ is also from $WLS$, and the square of a nonnegative functional from $WLS$ is in $WLS$, it suffices to prove that the functional
$$
G_j(h)=\left\{\int_0^T\left[\frac{j(s)}{\bar{\rho}\sigma\left({\cal K}h(s)\right)}-\frac{\rho}{\bar{\rho}}h(s)
\right]^2ds\right\}^{\frac{1}{2}}
$$
belongs to the class $WLS$. We have
$$
G_j(h)=\sup_{||q||_2\le 1}\int_0^T\left[\frac{j(s)}{\bar{\rho}\sigma\left({\cal K}h(s)\right)}-\frac{\rho}{\bar{\rho}}h(s)
\right]q(s)ds.
$$

The supremum of any family of functionals from $WLS$ is in $WLS$. Therefore, to finish the proof of Lemma \ref{L:dobb} it is enough to 
show that for any $j,q\in L^2[0,T]$, the functional 
$$
h\mapsto\int_0^T\left[\frac{j(s)}{\bar{\rho}\sigma\left({\cal K}h(s)\right)}-\frac{\rho}{\bar{\rho}}h(s)
\right]q(s)ds
$$
belongs to the class $WLS$. It is clear that the functional 
$h\mapsto-\frac{\rho}{\bar{\rho}}\int_0^Th(s)q(s)ds$ is weakly continuous, hence it is in $WLS$. It remains to analyze the functional
\begin{equation}
D_{j,q}(h)\mapsto\int_0^T\frac{j(s)q(s)}{\bar{\rho}\sigma\left({\cal K}h(s)\right)}ds.
\label{E:e}
\end{equation}
We will prove that this functional is weakly sequentially continuous. Suppose $h_k\rightarrow h$ weakly in $L^2[0,T]$. Then $\sup_k||h_k||_2<\infty$.
It follows from the restrictions on the kernel $K$ that the operator ${\cal K}$ is continuous from $L^2[0,T]$ into $\mathbb{C}[0,T]$. Therefore, it is weakly continuous, and hence ${\cal K}h_k(s)\mapsto{\cal K}h(s)$ for all $s\in[0,T]$. Moreover, we have
$$
\sup_{k\ge 1,s\in[0,T]}|{\cal K}h_k(s)|\le||{\cal K}||\sup_k||h_k||_2<\infty.
$$
It is assumed in Theorem \ref{T:2} that the volatility function $\sigma$ is strictly positive on $\mathbb{R}$. Therefore, it is bounded away from zero on any finite interval. Next, using (\ref{E:e}) and the dominated convergence theorem, we see that $D_{j,q}(h_k)\mapsto D_{j,q}(h)$ as $k\rightarrow\infty$. Hence the functional $D_{j,q}$ is weakly sequentially continuous. Finally, by taking into account the facts established above, it is easy to complete the proof of Lemma \ref{L:dobb}.

In the next lemma, we prove the continuity of the rate function $Q_T$ on the space $H^1_0[0,T]$.
\begin{lemma}\label{L:concon}
The functional $Q_T:H^1_0[0,T]\mapsto\mathbb{R}$ is continuous.
\end{lemma}

\it Proof. \rm The lower semi-continuity of the functional $Q_T$ in Lemma \ref{L:concon} follows from the fact that $Q_T$ is a rate function on $\mathbb{C}_0[0,T]$ and the Sobolev embedding $H^1_0[0,T]\subset\mathbb{C}_0[0,T]$. 

We will next prove the upper semi-continuity of $Q_T$ on $H^1_0[0,T]$. For every $f\in H^1_0[0,T]$, define the functional
${\cal D}_f:H^1_0[0,T]\mapsto\mathbb{R}$ by
$$
{\cal D}_f(g)=\int_0^T\left[\frac{\dot{g}(s)-\rho\sigma(\widehat{f}(s))\dot{f}(s)}
{\bar{\rho}\sigma(\widehat{f}(s))}\right]^2ds
+\int_0^T\dot{f}(s)^2ds.
$$
It is not hard to see that in order to complete the proof of Lemma \ref{L:concon}, it suffices to establish that for every 
$f\in H^1_0[0,T]$,
the functional ${\cal D}_f$ is continuous. It is clear that $\widehat{f}$ is a bounded continuous function on $[0,T]$. Therefore
there exist $\delta> 0$ and $M> 0$ such that
$M>\sigma(\widehat{f}(s))>\delta$ for all $s\in[0,T]$. Suppose $g_k\rightarrow g$ in $H^1_0[0,T]$. Then we have
\begin{align*}
&|{\cal D}_f(g)-{\cal D}_f(g_k)|\le\frac{1}{\delta^2(1-\rho^2)}\int_0^T|\dot{g}(s)-\dot{g}_k(s)|
|\dot{g}(s)+\dot{g}_k(s)-2\rho\sigma(\widehat{f}(s))\dot{f}(s)|ds \\
&\le\frac{1}{\delta^2(1-\rho^2)}||g-g_k||_{H^1_0[0,T]}\left(||g||_{H^1_0[0,T]}+\sup_k||g||_{H^1_0[0,T]}
+2\left\{\int_0^T\sigma(\widehat{f}(s))^2\dot{f}(s)^2ds\right\}^{\frac{1}{2}}\right) \\
&\le\le\frac{1}{\delta^2(1-\rho^2)}||g-g_k||_{H^1_0[0,T]}\left(||g||_{H^1_0[0,T]}+\sup_k||g||_{H^1_0[0,T]}
+2M||f||_{H^1_0[0,T]}\right).
\end{align*}
Now, it is clear that ${\cal D}_f(g_k)\rightarrow{\cal D}_f(g)$ as $k\rightarrow\infty$, and hence the functional ${\cal D}_f$
is continuous on the space $H^1_0[0,T]$. It follows that the functional $Q_T$ is upper semi-continuous since it can be represented as the infimum of a family of continuous on $H^1_0[0,T]$ functionals.

This completes the proof of Lemma \ref{L:concon}.

Our final goal in the present section is to apply Theorem \ref{T:2} to characterize the leading term in the asymptotic expansion of the exit probability from an open interval. Recall that we assumed $S_0^{\varepsilon, 0, H}=1$. Therefore, $X_0^{\varepsilon, 0, H}=0$. 
Let $U=(a,b)$ be an interval such that $0\in(a,b)$, and define the exit time from $U$ by
$
\tau^{\varepsilon}=\inf\left\{s\in(0,T]:X_s^{\varepsilon, 0, H}\notin U\right\}.
$
For every fixed $t\in(0,T]$, the exit time probability function $v_{\varepsilon}(t)$ is defined by
$
v_{\varepsilon}(t)=\mathbb{P}(\tau^{\varepsilon}\le t).
$
Set
$$
{\cal A}_t=\left\{f\in\mathbb{C}_0[0,T]:f(s)\notin U
\,\,\mbox{for some}\,\,s\in(0,t]\right\}.
$$
\begin{theorem}\label{T:33}
Under the conditions in Theorem \ref{T:2}, for every $t\in[0,T]$,
$$
\lim_{\varepsilon\rightarrow 0}\varepsilon^{2H}\log v_{\varepsilon}(t)=-\inf_{f\in{\cal A}_t}Q_T(f).
$$
\end{theorem}

\it Proof. \rm It is not hard to see that $\left\{\tau^{\varepsilon}\le t\right\}=\left\{X^{\varepsilon, 0, H}\in{\cal A}_t\right\}$.
We will next show that ${\cal A}_t$ is a set of continuity for $Q_T$. Indeed, the interior ${\cal A}^{\circ}_t$ of ${\cal A}_t$ 
consists of all the paths $f\in\mathbb{C}_0[0,T]$, for which there exists $s< t$ such that $f(s)\notin\bar{U}$. In addition, 
the boundary of ${\cal A}_t$ coincides with the set of all paths $f\in\mathbb{C}_0[0,T]$, which hit the boundary of $U$ before $t$ or at $s=t$, 
but never exit $\bar{U}$ before $t$. It is not hard to see that the set ${\cal A}^{\circ}_t\cup H_0^1[0,T]$ is dense in the set
$\bar{A}_t\cup H_0^1[0,T]$ in the topology of the space $H_0^1[0,T]$. Now, using Lemma \ref{L:concon}, it is easy to prove 
that the equality in (\ref{E:err}) holds for the set ${\cal A}_t$, and hence (\ref{E:errs}) is valid for ${\cal A}_t$.

The proof of Theorem \ref{T:33} is thus completed.

\section{Moderate deviations: $0<\beta< H$}\label{S:mdr}
In this section, we assume that $0<\beta< H$, and prove a sample path large deviation principle for the process $\varepsilon\mapsto X^{\varepsilon,\beta,H}$. We also obtain a similar result for the process $\varepsilon\mapsto X^{\varepsilon,\beta,H}_T$. It is not assumed 
in the present section that the Gaussian stochastic volatility model is of Volterra type.

The next statement is the main result of the present section.
\begin{theorem}\label{T:tet}
Let $0<\beta< H$, $\sigma(0)> 0$, and suppose the function $\sigma$ is locally $\omega$-continuous on $\mathbb{R}$ for some modulus of continuity $\omega$. Suppose also that $\widehat{B}$ is a nondegenerate continuous centered Gaussian process that is adapted to the filtration $\{\mathcal{F}_t\}_{0\le t\le T}$.
Then the process $\varepsilon\mapsto X^{\varepsilon,\beta,H}$ 
with state space $\mathbb{C}_0[0,T]$ satisfies the LDP with speed $\varepsilon^{2\beta-2H}$ and good rate function defined by
$$
\widetilde{I}_T(f)=
\begin{cases}
\frac{1}{2\sigma(0)^2}\int_0^T\dot{f}(t)^2dt, & f\in\mathbb{H}_0^1[0,T] \\
\infty, & f\in\mathbb{C}_0[0,T]\backslash\mathbb{H}_0^1[0,T].
\end{cases}
$$ 
\end{theorem}

\it Proof. \rm Set
\begin{equation}
\widehat{X}^{\varepsilon,\beta,H}_t=\varepsilon^{H-\beta}\int_0^t\sigma(\varepsilon^{H}\widehat{B}_s)dZ_s,\quad 0\le\varepsilon\le 1.
\label{E:widd}
\end{equation}
We will first prove that in the environment of Theorem \ref{T:tet}, the removal of the drift term does not affect the validity of the LDP. 
\begin{lemma}\label{L:first}
Under the conditions in Theorem \ref{T:tet}, the processes $\varepsilon\rightarrow\widehat{X}^{\varepsilon,\beta,H}$ and 
$\varepsilon\rightarrow X^{\varepsilon,\beta,H}$ with state space $\mathbb{C}_0[0,T]$ are exponentially equivalent.
\end{lemma}
\begin{remark}\label{R:ouo}
The definition of the exponential equivalence can be found in \cite{DZ}. In our case, the exponential equivalence means that
for every $y> 0$,
$$
\lim_{\varepsilon\rightarrow 0}\varepsilon^{2H-2\beta}
\log\mathbb{P}\left(||\widehat{X}^{\varepsilon,\beta,H}-X^{\varepsilon,\beta,H}||_{\mathbb{C}_0[0,T]}\ge y\right)=-\infty.
$$
\end{remark}

\it Proof of Lemma \ref{L:first}. \rm A statement similar to that in Lemma \ref{L:first} was obtained in a little different 
setting in Section 5 of \cite{G}. 
In our case, 
\begin{align*}
&\mathbb{P}\left(||\widehat{X}^{\varepsilon,\beta,H}-X^{\varepsilon,\beta,H}||_{\mathbb{C}_0[0,T]}\ge y\right)
=\mathbb{P}\left(\frac{1}{2}\varepsilon^{2H-2\beta}\int_0^T\sigma(\varepsilon^{H}\widehat{B}_s)^2ds\ge y\right), 
\end{align*}
and we can finish the proof of Lemma \ref{L:first}, using the same tools as in the proof in Section 5 of \cite{G}.

It follows from Lemma \ref{L:first} that the processes $\varepsilon\rightarrow\widehat{X}^{\varepsilon,\beta,H}$ and 
$\varepsilon\rightarrow X^{\varepsilon,\beta,H}$ satisfy the same large deviation principle (see \cite{DZ} for the proof of the 
fact that the exponential equivalence of two processes implies that they satisfy the same LDP). Hence, it suffices to prove 
Theorem \ref{T:tet} for the former process.
\begin{lemma}\label{L:summa}
Under the conditions in Theorem \ref{T:tet}, the process $\varepsilon\mapsto\widehat{X}^{\varepsilon,\beta,H}$ 
is exponentially equivalent to the process 
$\varepsilon\mapsto\widetilde{G}^{\varepsilon,\beta,H}:=\varepsilon^{H-\beta}\sigma(0)Z$.
\end{lemma}

\it Proof of Lemma \ref{L:summa}. \rm Let $\delta> 0$ and $0<\eta< 1$. For every $\varepsilon\in[0,1]$, set 
$$
M_t^{(\varepsilon)}=\int_0^t\left[\sigma(\varepsilon^H\widehat{B}_s)-\sigma(0)\right]dZ_s,\quad 0\le t\le T,
$$
and define a stopping time by
$
\xi_{\eta}^{(\varepsilon)}=\inf\left\{s\in[0,T]:\varepsilon^H|\widehat{B}_s|>\eta\right\}.
$
Then we have 
\begin{align}
&\mathbb{P}\left(||\widehat{X}^{\varepsilon,\beta,H}-\widetilde{G}^{\varepsilon,\beta,H}||_{\mathbb{C}_0[0,T]}>\delta\right)
=\mathbb{P}\left(\varepsilon^{H-\beta}\sup_{t\in[0,T]}
\left|M_t^{(\varepsilon)}\right|>\delta\right)
\nonumber \\
&\le\mathbb{P}\left(\varepsilon^{H-\beta}\sup_{t\in[0,\xi_{\eta}^{(\varepsilon)}]}
\left|M_t^{(\varepsilon)}\right|>\frac{\delta}{2}\right)
+\mathbb{P}\left(\xi_{\eta}^{(\varepsilon)}< T\right) \nonumber \\
&=J_1(\varepsilon,\delta,\eta)+J_2(\varepsilon,\delta,\eta).
\label{E:liul}
\end{align}

We will first estimate $J_1$. Set $\sigma_s^{(\varepsilon)}=\sigma(\varepsilon^H\widehat{B}_s)-\sigma(0)$.
Since the function $\sigma$ is locally $\omega$-continuous (see Definition \ref{D:moc}), 
\begin{equation}
|\sigma_s^{(\varepsilon)}|\le L(1)\omega(\eta)\quad\mbox{for all}\quad s\in\left[0,\xi_{\eta}^{(\varepsilon)}\right].
\label{E:sk}
\end{equation}

It is clear that since the process 
\begin{equation}
t\mapsto M(t\wedge\xi_{\eta}^{(\varepsilon)}),\quad t\in[0,T].
\label{E:marti1}
\end{equation} 
can be represented as a stochastic integral with a bounded integrand, it is a martingale.
Moreover, for $0<\varepsilon<\varepsilon_0$ and a fixed $\lambda> 0$, the stochastic exponential
$$
{\cal E}_t^{(\varepsilon)}=\exp\left\{\lambda\varepsilon^{H-\beta}\int_0^{t\wedge\xi_{\eta}^{(\varepsilon)}}
\sigma_s^{(\varepsilon)}dZ_s-\frac{1}{2}\lambda^2\varepsilon^{2H-2\beta}
\int_0^{t\wedge\xi_{\eta}^{(\varepsilon)}}
\left(\sigma_s^{(\varepsilon) }\right)^2ds\right\}
$$
is a martingale (use Novikov's condition). We will assume in the rest of the proof that 
$0<\varepsilon<\varepsilon_0$. 
It follows from (\ref{E:sk}) and the martingality condition formulated above
that
\begin{align}
&\mathbb{E}\left[\exp\left\{\lambda\varepsilon^{H-\beta}\int_0^{t\wedge\xi_{\eta}^{(\varepsilon)}}
\sigma_s^{(\varepsilon)}dZ_s\right\}\right]
=\mathbb{E}\left[{\cal E}_t^{(\varepsilon)}\exp\left\{\frac{1}{2}\lambda^2\varepsilon^{2H-2\beta}
\int_0^{t\wedge\xi_{\eta}^{(\varepsilon)}}
\left(\sigma_s^{(\varepsilon)}\right)^2ds\right\}\right]\nonumber \\
&\le\exp\left\{\frac{1}{2}T\lambda^2\varepsilon^{2H-2\beta}L(1)^2\omega(\eta)^2\right\}<\infty,
\label{E:nak1}
\end{align}
for all $t\in[0,T]$. Plugging $t=T$ into (\ref{E:nak1}), we get
\begin{equation}
\mathbb{E}\left[\exp\left\{\lambda\varepsilon^{H-\beta}\int_0^{\xi_{\eta}^{(\varepsilon)}}\sigma_s^{(\varepsilon)}dZ_s\right\}\right]
\le\exp\left\{\frac{1}{2}T\lambda^2\varepsilon^{2H-2\beta}L(1)^2\omega(\eta)^2\right\}.
\label{E:nak2}
\end{equation}

Since the process in (\ref{E:marti1}) is a martingale, the integrability condition in (\ref{E:nak1}) implies that the process 
$$
t\mapsto\exp\left\{\lambda\varepsilon^{H-\beta}\int_0^{t\wedge\xi_{\eta}^{(\varepsilon)}}\sigma_s^{(\varepsilon)}dZ_s\right\}
$$
is a positive submartingale (see Proposition 3.6 in \cite{KaS}). Next, using (\ref{E:nak2}) and the first submartingale inequality in 
\cite{KaS}, Theorem 3.8, we obtain
\begin{align*}
&\mathbb{P}\left(\sup_{t\in[0,\xi_{\eta}^{(\varepsilon)}]}
\exp\left\{\varepsilon^{H-\beta}\lambda\int_0^t\sigma_s^{(\varepsilon)}dZ_s\right\}> e^{\lambda\delta}\right) \\
&\le\exp\left\{\frac{1}{2}T\varepsilon^{2H-2\beta}\lambda^2
L(1)^2\omega(\eta)^{2}-\lambda\delta\right\}.
\end{align*}
Setting $\lambda=\frac{\delta}{T\varepsilon^{2H-2\beta}L(1)^2\omega(\eta)^{2}}$, we get from the previous inequality that
\begin{align}
&\mathbb{P}\left(\sup_{t\in[0,\xi_{\eta}^{(\varepsilon)}]}
\varepsilon^{H-\beta}\int_0^t\sigma_s^{(\varepsilon)}dZ_s>\delta\right) 
\le\exp\left\{-\frac{\delta^2}{2T\varepsilon^{2H-2\beta}L(1)^2\omega(\eta)^{2}}\right\}.
\label{E:five}
\end{align}

It is possible to replace the process $M$ by the process $-M$ in the reasoning above. This gives the following inequality that is similar to 
(\ref{E:five}): 
\begin{align}
&\mathbb{P}\left(\sup_{t\in[0,\xi_{\eta}^{(\varepsilon)}]}\left[-
\varepsilon^{H-\beta}\int_0^t\sigma_s^{(\varepsilon)}dZ_s\right]>\delta\right) 
\le\exp\left\{-\frac{\delta^2}{2T\varepsilon^{2H-2\beta}L(1)^2\omega(\eta)^{2}}\right\}.
\label{E:fiv}
\end{align}
It follows from (\ref{E:five}) and (\ref{E:fiv}) that
\begin{equation}
\mathbb{P}\left(\sup_{t\in[0,\xi_{\eta}^{(\varepsilon)}]}
\varepsilon^{H-\beta}\left|\int_0^t\sigma_s^{(\varepsilon)}dZ_s\right|>\delta\right)
\le 2\exp\left\{-\frac{\delta^2}{2T\varepsilon^{2H-2\beta}L(1)^2\omega(\eta)^{2}}\right\},
\label{E:hpo}
\end{equation}
for all $\delta> 0$ and $0<\eta< 1$. Then we have the following estimate for $J_1$ introduced in (\ref{E:liul}) 
$$
J_1(\varepsilon,\delta,\eta)\le 2\exp\left\{-\frac{\delta^2}{8T\varepsilon^{2H-2\beta}L(1)^2\omega(\eta)^{2}}\right\}
$$
and
\begin{equation}
\limsup_{\varepsilon\rightarrow 0}\varepsilon^{2H-2\beta}\log J_1(\varepsilon,\delta,\eta)
\le-\frac{\delta^2}{8TL(1)^2\omega(\eta)^{2}}.
\label{E:!}
\end{equation}

Our next goal is to estimate $J_2$ defined in (\ref{E:liul}). We have
\begin{equation}
J_2(\varepsilon,\delta,\eta)\le\mathbb{P}\left(\varepsilon^H\sup_{s\in[0,T]}|\widehat{B}_s|>\eta\right),
\label{E:shcu}
\end{equation}
for all $\varepsilon\in(0,T]$, $\delta> 0$, and $\eta\in(0,1)$. Using the large deviation principle for the maximum of a Gaussian process 
(see, e.g., (8.5) in \cite{L}), we can show that there exist constants $C_1> 0$ and $y_0> 0$ such that
\begin{equation}
\mathbb{P}\left(\sup_{t\in[0,T]}|\widehat{B}_t|> y\right)\le e^{-C_1 y^2}
\label{E:addik}
\end{equation}
for all $y> y_0$. Next, taking into account (\ref{E:shcu}) and (\ref{E:addik}), we obtain
\begin{equation}
\limsup_{\varepsilon\rightarrow 0}\varepsilon^{2H-2\beta}\log J_2(\varepsilon,\delta,\eta)=-\infty.
\label{E:!!}
\end{equation}

Finally, combining (\ref{E:liul}), (\ref{E:!}), and (\ref{E:!!}), and using the inequality 
$$
\log(a+b)\le\max\{\log(2a),\log(2b)\},\quad a> 0,\quad b> 0,
$$ 
we can prove that 
$$
\lim_{\varepsilon\rightarrow 0}\varepsilon^{2H-2\beta}\log\mathbb{P}\left(||\widehat{X}^{\varepsilon,\beta,H}-\widetilde{G}^{\varepsilon,\beta,H}||_{\mathbb{C}_0[0,T]}>\delta\right)=-\infty,
$$
for all $\delta> 0$.

The proof of Lemma \ref{L:summa} is thus completed.

To finish the proof of Theorem \ref{T:tet}, we observe that by Schilder's theorem (see \cite{DZ}), the process $\widetilde{G}^{\varepsilon,\beta,H}$ satisfies the LDP in the formulation of Theorem \ref{T:tet}. Next, using the exponential equivalence in Lemmas \ref{L:first}
and \ref{L:summa}, we see that the same LDP holds for the process $\widehat{X}^{\varepsilon,\beta,H}$.

This completes the proof of Theorem \ref{T:tet}.
\begin{corollary}\label{C:corio}
Under the restrictions in Theorem \ref{T:tet}, the process $\varepsilon\mapsto X^{\varepsilon,\beta,H}_T$ 
with state space $\mathbb{R}$ satisfies the LDP with speed $\varepsilon^{2\beta-2H}$ and good rate function defined by
$$
\widehat{I}_T(x)=\frac{x^2}{2T\sigma(0)^2},\quad x\in\mathbb{R}.
$$
\end{corollary}

Corollary \ref{C:corio} can be derived from Theorem \ref{T:tet}. Indeed, let $A$ be a Borel subset of $\mathbb{R}$, and consider the Borel subset $\widetilde{A}$ of $\mathbb{C}_0$ consisting of $f\in\mathbb{C}_0[0,T]$
such that $f(T)\in A$. Then, it is not hard to prove the LDP-estimates in Corollary \ref{C:corio} 
for the set $A$, by applying the LDP-estimates in Theorem \ref{T:tet} to the set $\widetilde{A}$.
\begin{remark}
The large deviation and moderate deviation results obtained in Theorem
\ref{T:1} and Corollary \ref{C:corio}, and also the fact that the rate function $I_T$ is nondecreasing on $[0,\infty)$ (see \cite{G}), imply the following tail estimates:
\begin{equation}
\lim_{\varepsilon\downarrow 0}\varepsilon^{2H-2\beta}\log\mathbb{P}\left(X_T^{\varepsilon,\beta,H}\ge x\right)
=
\begin{cases}
-I_T(x), &\mbox{if}\quad \beta=0 \\
-\frac{x^2}{2T\sigma(0)^2}, &\mbox{if}\quad 0<\beta< H.
\end{cases} 
\label{E:discussi}
\end{equation}
\end{remark}

Our next goal is to discuss relations between small-time and small-noise LDPs for self-similar volatility processes.
\begin{definition}\label{D:ssi}
Let $0< H< 1$. The process $\widehat{B}_t$, $0\le t\le T$, is called $H$-self-similar if for every
$\varepsilon\in(0,1]$, $\widehat{B}_{\varepsilon t}=\varepsilon^H\widehat{B}_t$, $0\le t\le T$, in law. 
\end{definition}
Fractional Brownian motion $B^H$ and the Riemann-Liouville fractional Brownian motion $R^H$ are $H$-self-similar, 
while the fractional Ornstein-Uhlenbeck process $U^H$ is not. 

Suppose the volatility process $\widehat{B}$ is $H$-self-similar. 
Then, we can pass from small-noise LDP and MDP
in Theorem \ref{T:1} and Corollary \ref{C:corio} 
to small-time LDP and MDP, by making the following observation (such methods are well-known). Set
$$
Z_t^{\varepsilon,H}=-\frac{\varepsilon}{2}\int_0^t\sigma(\varepsilon^H\widehat{B}_s)^2ds+\sqrt{\varepsilon}\int_0^t
\sigma(\varepsilon^H\widehat{B}_s)dZ_s.
$$
It is not hard to see that under the self-similarity condition for $\widehat{B}$, for every $\varepsilon\in(0,1]$ the equality
$X_{\varepsilon t}=Z_t^{\varepsilon,H}$, $0\le t\le T$, holds in law. Here the process $X$ is defined by (\ref{E:poz}). Then 
$$
\varepsilon^{H-\beta-\frac{1}{2}}X_{\varepsilon t}=-\frac{1}{2}\varepsilon^{H-\beta+\frac{1}{2}}
\int_0^t\sigma(\varepsilon^H\widehat{B}_s)^2ds+\widehat{X}_t^{\varepsilon,\beta,H},
$$
for all $t\in[0,T]$, $\varepsilon\in(0,1]$, $H\in(0,1)$, and $\beta\in[0,H)$. The process $\widehat{X}_t^{\varepsilon,\beta,H}$ 
in the previous equality is defined in (\ref{E:intl}) and (\ref{E:widd}). Next, replacing $t$ by $T$ and $\varepsilon$
by $t$, we obtain
\begin{equation}
t^{H-\beta-\frac{1}{2}}X_{tT}=-\frac{1}{2}t^{H-\beta+\frac{1}{2}}\int_0^T\sigma(t^H\widehat{B}_s)^2ds+\widehat{X}_T^{t,\beta,H},
\label{E:triu}
\end{equation}
for all $t\in[0,1]$. For $\beta=0$ the process $t\mapsto X_T^{t,\beta,H}$ satisfies the LDP in Theorem \ref{T:1}, while for $\beta\in(0,H)$
it satisfies the MDP in Corollary \ref{C:corio}.
Now, replacing the drift term $-\frac{1}{2}t^{2H-2\beta}\int_0^T\sigma(t^H\widehat{B}_s)^2ds$ in the process 
$t\mapsto X_T^{t,\beta,H}$ by a new drift term 
$$
-\frac{1}{2}t^{H-\beta+\frac{1}{2}}\int_0^T\sigma(t^H\widehat{B}_s)^2ds,
$$ 
and using (\ref{E:triu}), we see that the process on the left-hand side of (\ref{E:triu}) satisfies the LDP in Theorem \ref{T:1} for $\beta=0$,
and the MDP in Corollary \ref{C:corio} for $\beta\in(0,H)$. The possibility of drift replacement can be justified using the ideas employed in Section 5 of \cite{G}. The previous reasoning shows how to obtain small-time large and moderate deviation principles from the small-noise ones.
For $\beta=0$, a small-time analogue of the LDP in Theorem \ref{T:1} was obtained in \cite{G}, Theorem 18.
\section{Central limit regime: $\beta=H$}\label{S:clr}
We will next describe what happens if $\beta=H$. Recall that in LDP and MDP regimes, we can ignore drift terms. For $\beta=H$, 
this is no more the case, and drift terms have to be taken into account. In the rest of the paper, the symbol $\bar{{\cal N}}$ will 
stand for the standard normal complementary cumulative distribution function defined by
$$
\bar{{\cal N}}(z)=\frac{1}{\sqrt{2\pi}}\int_z^{\infty}\exp\left\{-\frac{u^2}{2}\right\}du,\quad z\in\mathbb{R}.
$$

Let us assume that the restrictions on the function $\sigma$ imposed in 
Theorem \ref{T:tet} hold. We have 
$$
X^{\varepsilon,H,H}_t=-\frac{1}{2}\int_0^t\sigma(\varepsilon^{H}\widehat{B}_s)^2ds+\int_0^t\sigma(\varepsilon^{H}\widehat{B}_s)dZ_s,
\quad 0\le t\le T.
$$
If $\beta=H$, then the expression on the left-hand side of (\ref{E:discussi}) has the following form:
\begin{equation}
L(x)=\lim_{\varepsilon\downarrow 0}\log\mathbb{P}\left(X_T^{\varepsilon,H,H}\ge x\right),\quad x> 0.
\label{E:limi}
\end{equation}
It will be shown below that the limit in (\ref{E:limi}) exists for every $x> 0$, and its value will be computed.

We will first study the behavior of the
process 
$\varepsilon\mapsto X^{\varepsilon,H,H}$ on the path space. Set 
\begin{equation}
U_t=-\frac{1}{2}t\sigma(0)^2+\sigma(0)Z_t,\quad t\in[0,T].
\label{E:eq}
\end{equation}
\begin{theorem}\label{T:theor}
Under the restrictions on the function $\sigma$ and the process $\widehat{B}$ imposed in 
Theorem \ref{T:tet}, the following formula holds for all $y> 0$:
$$
\lim_{\varepsilon\rightarrow 0}\mathbb{P}\left(||X^{\varepsilon,H,H}-U||_{\mathbb{C}_0[0,T]}\ge y\right)=0.
$$
\end{theorem}

\it Proof. \rm For every $y> 0$, 
\begin{align}
&\mathbb{P}\left(||X^{\varepsilon,H,H}-U||_{\mathbb{C}_0[0,T]}\ge y\right)\le
\mathbb{P}\left(\sup_{t\in[0,T]}\left|\int_0^t\left[\sigma(0)^2-\sigma(\varepsilon^{H}\widehat{B}_s)^2\right]ds\right|\ge y\right) 
\nonumber \\
&\quad+\mathbb{P}\left(\sup_{t\in[0,T]}\left|\int_0^t\left[\sigma(\varepsilon^{H}\widehat{B}_s)-\sigma(0)\right]dZ_s\right|
\ge\frac{y}{2}\right) \nonumber \\
&=L_1(\varepsilon,y)+L_2(\varepsilon,y).
\label{E:er}
\end{align}

We will first show that
\begin{equation}
\lim_{\varepsilon\rightarrow 0}L_2(\varepsilon,y)=0.
\label{E:sis}
\end{equation}
To prove the equality in (\ref{E:sis}), we employ the methods used in the proof of Lemma \ref{L:summa}. Analyzing the proof preceding 
(\ref{E:hpo}), we see that the estimate in (\ref{E:hpo}) holds for $\beta=H$ too. This gives
$$
\mathbb{P}\left(\sup_{t\in[0,\xi_{\eta}^{(\varepsilon)}]}
\left|\int_0^t\sigma_s^{(\varepsilon)}dZ_s\right|>\delta\right)
\le 2\exp\left\{-\frac{\delta^2}{2L(1)^2\omega(\eta)^{2}}\right\}.
$$
Now, it is not hard to see how to prove (\ref{E:sis}) using (\ref{E:shcu}) and (\ref{E:addik}).

Our next goal is to show that
\begin{equation}
\lim_{\varepsilon\rightarrow 0}L_1(\varepsilon,y)=0.
\label{E:si}
\end{equation}
For all $\eta\in(0,1)$, we have
\begin{align}
&L_1(\varepsilon,y)\le\mathbb{P}\left(\sup_{t\in[0,\xi_{\eta}^{(\varepsilon)}]}
\left|\int_0^t\left[\sigma(0)^2-\sigma(\varepsilon^{H}\widehat{B}_s)^2\right]ds\right|\ge\frac{y}{2}\right)
+\mathbb{P}\left(\xi_{\eta}^{(\varepsilon)}< T\right) \nonumber \\
&\le\mathbb{P}\left(\sup_{t\in[0,\xi_{\eta}^{(\varepsilon)}]}
\int_0^t\left|\sigma(0)-\sigma(\varepsilon^{H}\widehat{B}_s)\right|
\left(\sigma(0)+\sigma(\varepsilon^{H}\widehat{B}_s)\right)ds\ge\frac{y}{2}\right)
+\mathbb{P}\left(\varepsilon^H\sup_{s\in[0,T]}|\widehat{B}_s|>\eta\right) \nonumber \\
&\le\mathbb{P}\left(2TL(1)\omega(\eta)\sup_{0\le u\le 1}[\sigma(u)]\ge\frac{y}{2}\right)+
\mathbb{P}\left(\varepsilon^H\sup_{s\in[0,T]}|\widehat{B}_s|>\eta\right).
\label{E:kon1}
\end{align}

For a fixed $y> 0$ and $\eta$ small enough, the first term on the last line in (\ref{E:kon1}) is equal to zero, since
$\omega(\eta)\rightarrow 0$ as $\eta\rightarrow 0$. Moreover, for a fixed $\eta\in(0,1)$, we have
$$
\lim_{\varepsilon\rightarrow 0}\mathbb{P}\left(\varepsilon^H\sup_{s\in[0,T]}|\widehat{B}_s|>\eta\right)=0.
$$
The previous equality can be obtained using (\ref{E:addik}). Now, it is not hard to see that (\ref{E:kon1}) implies (\ref{E:sis}).
Finally, it is clear that Theorem \ref{T:theor} follows from (\ref{E:er}), (\ref{E:sis}), and (\ref{E:si}).

The next result answers the following question asked by Barbara Pacchiarotti: What estimates for the  
function
$$
D_{H,T}(x)=\lim_{\varepsilon\rightarrow 0}\mathbb{P}\left(\sup_{t\in[0,T]}X^{\varepsilon,H,H}_t>x\right)
$$
can one derive using Theorem \ref{T:theor}? We will only deal with the right tail.
\begin{theorem}\label{T:rig}
Let $H> 0$ and $x> 0$. Then
\begin{align}
D_{H,T}(x)&=\Phi\left(\frac{x}{\sigma(0)\sqrt{T}}+\frac{\sigma(0)\sqrt{T}}
{2}\right) \nonumber \\
&\quad+\exp\{-x\}\left[1-\Phi\left(\frac{x}{\sigma(0)\sqrt{T}}+\frac{\sigma(0)\sqrt{T}}
{2}\right)\right].
\label{E:rex0}
\end{align}
where the symbol $\Phi$ stands for the standard normal cumulative distribution function.
\end{theorem}

\it Proof. \rm Fix $\delta$ with $0<\delta< 1$. Then we have
\begin{align*}
&\mathbb{P}\left(\sup_{t\in[0,T]}X^{\varepsilon,H,H}_t>x\right)\le\mathbb{P}\left(||X^{\varepsilon,H,H}-U||
_{\mathbb{C}_0[0,T]}>(1-\delta)x\right)
+\mathbb{P}\left(\sup_{t\in[0,T]}U_t>\delta x\right),
\end{align*}
where $U$ is defined by (\ref{E:eq}). It follows from Theorem \ref{T:theor} that
\begin{align}
&\limsup_{\varepsilon\rightarrow 0}
\mathbb{P}\left(\sup_{t\in[0,T]}X^{\varepsilon,H,H}_t>x\right) \nonumber \\
&\le\limsup_{\varepsilon\rightarrow 0}\mathbb{P}\left(||X^{\varepsilon,H,H}-U||
_{\mathbb{C}_0[0,T]}>(1-\delta)x\right)
+\mathbb{P}\left(\sup_{t\in[0,T]}U_t>\delta x\right) \nonumber \\
&=\mathbb{P}\left(\sup_{t\in[0,T]}U_t>\delta x\right)=\mathbb{P}\left(\sup_{t\in[0,T]}\left(-\frac{1}{2}t\sigma(0)+Z_t\right)
>\frac{\delta x}{\sigma(0)}\right).
\label{E:rex1}
\end{align}

The distribution of the maximum of Brownian motion with drift is known. The following formula holds for every $y> 0$ and $\mu\in\mathbb{R}$:
\begin{equation}
\mathbb{P}\left(\sup_{t\in[0,T]}\left(\mu t+Z_t\right)>y\right)=\Phi\left(\frac{y-\mu T}{\sqrt{T}}\right)
+\exp\{2\mu y\}\left[1-\Phi\left(\frac{y-\mu T}{\sqrt{T}}\right)\right]
\label{E:rex3}
\end{equation}
(see, e.g., \cite{ALC}). Therefore, (\ref{E:rex1}) implies that
\begin{align*}
\limsup_{\varepsilon\rightarrow 0}
\mathbb{P}\left(\sup_{t\in[0,T]}X^{\varepsilon,H,H}_t>x\right)&\le\Phi\left(\frac{\delta x}{\sigma(0)\sqrt{T}}+\frac{\sigma(0)\sqrt{T}}
{2}\right) \\
&\quad+\exp\{-\delta x\}\left[1-\Phi\left(\frac{\delta x}{\sigma(0)\sqrt{T}}+\frac{\sigma(0)\sqrt{T}}
{2}\right)\right].
\end{align*}
Next, by letting $\delta\rightarrow 1$, we obtain
\begin{align}
\limsup_{\varepsilon\rightarrow 0}
\mathbb{P}\left(\sup_{t\in[0,T]}X^{\varepsilon,H,H}_t>x\right)&\le\Phi\left(\frac{x}{\sigma(0)\sqrt{T}}+\frac{\sigma(0)\sqrt{T}}
{2}\right) \nonumber \\
&\quad+\exp\{-x\}\left[1-\Phi\left(\frac{x}{\sigma(0)\sqrt{T}}+\frac{\sigma(0)\sqrt{T}}
{2}\right)\right].
\label{E:rex4}
\end{align}

Our next goal is to obtain a lower estimate. We have
\begin{align*}
&\mathbb{P}\left(\sup_{t\in[0,T]}X^{\varepsilon,H,H}_t>x\right)\ge\mathbb{P}\left(\sup_{t\in[0,T]}U_t>(1+\delta)x\right) 
-\mathbb{P}\left(||X^{\varepsilon,H,H}-U||
_{\mathbb{C}_0[0,T]}>\delta x\right)
\end{align*}
and
\begin{align}
&\liminf_{\varepsilon\rightarrow 0}\mathbb{P}\left(\sup_{t\in[0,T]}X^{\varepsilon,H,H}_t>x\right) \nonumber \\
&\ge
\mathbb{P}\left(\sup_{t\in[0,T]}U_t>(1+\delta)x\right)-\limsup_{\varepsilon\rightarrow 0}\mathbb{P}\left(||X^{\varepsilon,H,H}-U||
_{\mathbb{C}_0[0,T]}>\delta x\right) \nonumber \\
&=\mathbb{P}\left(\sup_{t\in[0,T]}U_t>(1+\delta)x\right)=\mathbb{P}\left(\sup_{t\in[0,T]}\left(-\frac{1}{2}t\sigma(0)+Z_t\right)
>\frac{(1+\delta)x}{\sigma(0)}\right).
\label{E:rex5}
\end{align}
Now, using (\ref{E:rex5}), (\ref{E:rex3}), and letting $\delta\rightarrow 0$, we obtain
\begin{align}
\liminf_{\varepsilon\rightarrow 0}\mathbb{P}\left(\sup_{t\in[0,T]}X^{\varepsilon,H,H}_t>x\right)&\ge 
\Phi\left(\frac{x}{\sigma(0)\sqrt{T}}+\frac{\sigma(0)\sqrt{T}}
{2}\right) \nonumber \\
&\quad+\exp\{-x\}\left[1-\Phi\left(\frac{x}{\sigma(0)\sqrt{T}}+\frac{\sigma(0)\sqrt{T}}
{2}\right)\right]. 
\label{E:rex6}
\end{align}

Finally, it is clear that (\ref{E:rex0}) follows from (\ref{E:rex4}) and (\ref{E:rex6}).

This completes the proof of Theorem \ref{T:rig}.

The next statement is a corollary of Theorem \ref{T:theor}.
\begin{theorem}\label{T:clr}
Under the restrictions on the function $\sigma$ imposed in 
Theorem \ref{T:tet}, the following formula is valid:
$$
\lim_{\varepsilon\downarrow 0}\mathbb{P}\left(X_T^{\varepsilon,H,H}\ge x\right)=
\bar{{\cal N}}\left(\frac{x}{\sqrt{T}\sigma(0)}+\frac{1}{2}\sqrt{T}\sigma(0)\right).
$$
Therefore the limit in (\ref{E:limi}) exists for every $x> 0$, and moreover 
\begin{align}
&L(x)=
\log\bar{{\cal N}}\left(\frac{x}{\sqrt{T}\sigma(0)}+\frac{1}{2}\sqrt{T}\sigma(0)\right).
\label{E:disco}
\end{align}
\end{theorem}

\it Proof. \rm
By Theorem \ref{T:theor}, the process $X_T^{\varepsilon,H,H}$ converges in probability as $\varepsilon\downarrow 0$ to the random variable
$-\frac{1}{2}T\sigma(0)^2+\sigma(0)Z_T$. It is known that convergence in probability implies convergence in distribution. Since for every 
$x> 0$, the set $[x,\infty)$ is a set of continuity of the distribution of $Z_T$, we have
\begin{align}
\lim_{\varepsilon\downarrow 0}\mathbb{P}\left(X_T^{\varepsilon,H,H}\ge x\right)&=\frac{1}{\sqrt{2\pi}\sqrt{T}\sigma(0)}\int_x^{\infty}
\exp\left\{-\frac{1}{2T\sigma(0)^2}\left(r+\frac{1}{2}T\sigma(0)^2\right)^2\right\}dr \nonumber \\
&=\bar{{\cal N}}\left(\frac{x}{\sqrt{T}\sigma(0)}+\frac{1}{2}\sqrt{T}\sigma(0)\right).
\label{E:drisk}
\end{align}
Now it is clear that (\ref{E:disco}) follows from (\ref{E:drisk}).

This completes the proof of Theorem \ref{T:clr}.
\begin{remark}\label{R:uz1}
In the case where $\beta=H$, one can consider the function
$$
-L_T(x)=-\log\bar{{\cal N}}\left(\frac{x}{\sqrt{T}\sigma(0)}+\frac{1}{2}\sqrt{T}\sigma(0)\right),\quad x> 0,
$$
as a replacement for the rate function $I_T$ in the large deviation principle in Theorem \ref{T:1}, or the rate function 
$\widehat{I}_T(x)=\frac{1}{2\sqrt{T}\sigma(0)^2}x^2$ in the moderate deviation principle in Corollary \ref{C:corio}. However,
for $\beta=H$, the corresponding moderate deviation principle is degenerated since in this case the speed $\varepsilon^{2H-2\beta}$ is identically equal to one.
\end{remark}
\begin{remark}\label{R:di}
If $\beta\rightarrow 0$, then the rate function in the MDP regime in Corollary \ref{C:corio} does not tend to the rate function in the LDP regime in Theorem \ref{T:1}. This discontinuity disappears for small $x> 0$, if we tolerate an $O(x^3)$-approximation. Indeed for $\beta=0$, the following asymptotic expansion was established in \cite{BFGHS} under a stronger smoothness restriction on the volatility function:
$$
I_T(x)=\frac{x^2}{2T\sigma(0)^2}+O(x^3)
$$
as $x\rightarrow 0$ (actually, more terms in the Taylor expansion above were found in \cite{BFGHS}). Note that there is also a discontinuity in the asymptotic formulas at $\beta=H$. One of the reasons for the above-mentioned discontinuities is that it is in general not possible to pass to the limit with respect to an extra parameter in asymptotic formulas.
\end{remark}

\section{Asymptotic behavior of small-noise call pricing functions in mixed regimes}\label{S:varre}
In this section, we study the asymptotic behavior of small-noise call pricing functions in mixed regimes. 
The following small-noise call pricing functions will be considered:
\begin{align*}
&C^{\beta,H,T}(\varepsilon,x\varepsilon^{\alpha})=\mathbb{E}\left[\left(S_T^{\varepsilon,\beta,H}
-\exp\left\{x\varepsilon^{\alpha}\right\}\right)^{+}
\right].
\end{align*}
The parameters appearing in the previous definition satisfy $x> 0$, $H> 0$, $\beta\le H$, $\alpha\ge 0$, and
$0\le\alpha+\beta\le H$. Note that the parameter $\beta$ may take negative values. 
In the previous formula, the maturity is parametrized by $\varepsilon$, while the log-strike follows the path 
$\varepsilon\mapsto x\varepsilon^{\alpha}$ (see \cite{GL} for the discussion of various parametrizations of the call).

In the present section, we deal with the case where the volatility function satisfies
the linear growth condition. What happens when the volatility function grows at infinity faster than linearly is discussed 
in the next section.
\begin{definition}\label{D:lingr}
It is said that the linear growth condition holds for the function $\sigma$ if there exist constants $c_1> 0$ and 
$c_2> 0$ such that 
$\sigma(x)^2\le c_1+c_2x^2$ for all $x\ge 0$.
\end{definition}

It is known that if the linear growth condition holds for the function $\sigma$, then the asset price process $S$ in the model described by
(\ref{E:DD}) is a martingale,  and hence $\mathbb{P}$ is a risk-neutral measure (see, e.g., \cite{FZ,G}). The process $S$ can be a martingale even for more rapidly growing functions $\sigma$. For example, it was established in 
\cite{J} that for the Scott model (see \cite{Sc}), where $\sigma(x)=e^x$ and $\widehat{B}$ is the classical Ornstein-Uhlenbeck process, the asset price process $S$ is a martingale if and only if $-1<\rho\le 0$. A similar result for general Volterra type Gaussian models was obtained in a recent preprint \cite{Gassiat}. 

In the next assertion, asymptotic formulas for call pricing functions are derived from large deviation principles.
\begin{theorem}\label{T:terl}
Suppose the volatility function $\sigma$ satisfies the linear growth condition. Then the following are true: \\
(i)\,\,Assume that the conditions in Theorem \ref{T:1} hold, and let $\alpha+\beta=0$. Then
$$
\lim_{\varepsilon\downarrow 0}\varepsilon^{2H}
\log C^{\beta,H,T}(\varepsilon,x\varepsilon^{\alpha})=-I_T(x).
$$
(ii)\,\,Assume that the conditions in Corollary \ref{C:corio} hold, and let $0<\alpha+\beta< H$. Then
$$
\lim_{\varepsilon\downarrow 0}\varepsilon^{2H-2\alpha-2\beta}
\log C^{\beta,H,T}(\varepsilon,x\varepsilon^{\alpha})=-\frac{x^2}{2T\sigma(0)^2}.
$$
\end{theorem}

\it Proof. \rm The methods allowing to derive estimates for call pricing functions from large deviation principles under the linear growth condition are well-known (see, e.g., \cite{FJ,FZ,G,Ph}). We will only sketch the proof of the upper estimate in part (i) of Theorem \ref{T:terl}.  Let us start with tail estimates. It follows from (\ref{E:eq1}) and the possibility of removing the drift terms in the case where
$\alpha+\beta\neq H$ mentioned above that
\begin{align}
\lim_{\varepsilon\downarrow 0}\varepsilon^{2H-2\alpha-2\beta}
\log\mathbb{P}\left(X_T^{\varepsilon,\beta,H}\ge x\varepsilon^{\alpha}\right)&=
\lim_{\varepsilon\downarrow 0}\varepsilon^{2H-2\alpha-2\beta}\log\mathbb{P}\left(\varepsilon^{-\alpha}
X_T^{\varepsilon,\beta,H}\ge x\right) \nonumber \\
&=\lim_{\varepsilon\downarrow 0}\varepsilon^{2H-2\alpha-2\beta}
\log\mathbb{P}\left(X_T^{\varepsilon,\alpha+\beta,H}\ge x\right).
\label{E:d1}
\end{align}
If $\alpha+\beta=0$, then Theorem \ref{T:1} and the equality in (\ref{E:d1}) imply that
$$
\lim_{\varepsilon\downarrow 0}\varepsilon^{2H}
\log\mathbb{P}\left(X_T^{\varepsilon,\beta,H}\ge x\varepsilon^{\alpha}\right)=-I_T(x)
$$
(to prove a similar equality in the case where $0<\alpha+\beta< H$ we use Corollary \ref{C:corio} and (\ref{E:d1})).

Let $p> 1$ and $q> 1$ be such that $\frac{1}{p}+\frac{1}{q}=1$. Then
$$
C^{\beta,H,T}(\varepsilon,x\varepsilon^{\alpha})\le\left\{\mathbb{E}\left[\left|S_T^{\varepsilon,\beta,H}\right|^p\right]\right\}
^{\frac{1}{p}}
\left\{\mathbb{P}\left(X_T^{\varepsilon,\beta,H}> x\varepsilon^{\alpha}\right)\right\}^{\frac{1}{q}}.
$$
It can be seen from the previous estimate that
\begin{align*}
&\limsup_{\varepsilon\downarrow 0}\varepsilon^{2H-2\alpha-2\beta}
\log C^{\beta,H,T}(\varepsilon,x\varepsilon^{\alpha})
\le\frac{1}{p}\limsup_{\varepsilon\downarrow 0}\varepsilon^{2H-2\alpha-2\beta}
\log\mathbb{E}\left[\left|S_T^{\varepsilon,\beta,H}\right|^p\right]-\frac{1}{q}I_T(x). 
\end{align*}

The rest of the proof of the upper estimate in part (i) of Theorem \ref{T:terl} follows the proof of a similar estimate in 
Corollary 31 in \cite{G} (formula (81) in \cite{G}). By reasoning as in the latter proof we can establish that 
for all $\varepsilon\in(0,1]$, $p> 1$, and $\beta\in[0,H)$,
\begin{equation}
\mathbb{E}\left[\left|S_T^{\varepsilon,\beta,H}\right|^p\right]\le s_0^p
\left\{\mathbb{E}\left[\exp\left\{\left(2p^2-p\right)\varepsilon^{2H-2\beta}
\int_0^T\sigma(\varepsilon^H\widehat{B}_s)^2ds\right\}\right]\right\}^{\frac{1}{2}}.
\label{E:rt1}
\end{equation}
In the proof of (\ref{E:rt1}), the assumption that the asset price process $S$ is a martingale is used. 

It was shown in \cite{G} that
there exists $\delta_1> 0$ such that
\begin{equation}
\mathbb{E}\left[\exp\left\{\delta_1\int_0^T\widehat{B}_s^2ds\right\}\right]<\infty.
\label{E:sti}
\end{equation}
Next, using the linear growth condition for $\sigma$, we prove that the inequality in (\ref{E:sti}) 
implies the existence of $\varepsilon_0> 0$ and $\delta_2> 0$ for which
\begin{equation}
\sup_{\varepsilon\in(0,\varepsilon_0]}\mathbb{E}\left[\exp\left\{\delta_2\int_0^T\sigma(\varepsilon^H\widehat{B}_s)^2
ds\right\}\right]<\infty.
\label{E:stihi}
\end{equation}
Now, we can finish the proof of part (i) of Theorem \ref{T:terl}
exactly as in \cite{G}, Corollary 31. Note that the following fact is needed here. Since the linear growth condition holds, the 
stochastic exponential $t\mapsto S_t^{\varepsilon,\beta,H}$ defined in (\ref{E:DDES}) is a martingale for every fixed $\varepsilon$, and therefore
$
1=\mathbb{E}\left[S_T^{\varepsilon,\beta,H}\right]\le\mathbb{E}\left[|S_T^{\varepsilon,\beta,H}|^p\right]
$
for all $\varepsilon\in(0,1]$ and $p>1$.

This completes a short sketch of the proof of part (i) of Theorem \ref{T:terl}.
\begin{remark}\label{R:ooo}
It is interesting that if the volatility function $\sigma$ grows at infinity a little faster than linearly, then the inequality
in (\ref{E:stihi}) may fail (see Theorem \ref{T:confit}). This means that special methods of establishing the upper LDP and MDP call price estimates, which are based on the finiteness of the exponential moments of the integrated variance
can not be employed under even a slightly weaker than the linear growth restriction on the volatility function $\sigma$ 
(for more information see the discussion in Section \ref{S:revisit}). 
\end{remark}

It remains to characterize the asymptotic behavior of the call pricing function in the regime where $\alpha+\beta=H$. 
We will first restrict ourselves to the case where $\alpha=0$ and $\beta=H$.
\begin{theorem}\label{T:calli}
Suppose $\alpha=0$ and
$\beta=H$. Suppose also that the conditions in Corollary \ref{C:corio} are valid, and the function $\sigma$ satisfies the linear growth condition. 
Then the following formula holds:
\begin{align}
&\lim_{\varepsilon\downarrow 0} C^{H,H,T}(\varepsilon,x)=
\int_{x}^{\infty}e^y\bar{{\cal N}}\left(\frac{y}{\sqrt{T}\sigma(0)}+\frac{\sqrt{T}\sigma(0)}{2}\right)dy.
\label{E:integ}
\end{align}
\end{theorem}
\begin{remark}\label{R:oio}
The formula in (\ref{E:integ}) can be rewritten as follows:
\begin{align}
&\lim_{\varepsilon\downarrow 0}C^{H,H,T}(\varepsilon,x)=C_{-}(x,\sqrt{T}\sigma(0)),
\label{E:integro}
\end{align}
where the symbol $C_{-}(k,\nu)$ stands for the call price in the Black-Scholes model as a function of the log strike $k\ge 0$ and the dimensionless
implied volatility $\nu$ (see the definition in formula (3.1) in \cite{GL}). We leave the proof of the fact that the formulas in 
(\ref{E:integ}) and (\ref{E:integro}) are the same as an exercise for the interested reader. It follows from \cite{GL} (see 
the second equality in formula (3.1) and formula
(3.3) in \cite{GL}) that for every fixed $k$, $C_{-}$ is a strictly increasing function of $\nu$. 
\end{remark}

\it Proof of Theorem \ref{T:calli}. \rm Set
\begin{align}
&P_{\varepsilon}^{\alpha,H,T}(x)=
\mathbb{P}\left(-\frac{1}{2}\varepsilon^{\alpha}\int_0^T\sigma(\varepsilon^{H}\widehat{B}_s)^2ds+
\int_0^T\sigma(\varepsilon^{H}\widehat{B}_s)dZ_s\ge x\right).
\label{E:vr}
\end{align}
It is not hard to see that 
\begin{align}
C^{H,H,T}(\varepsilon,x)&=\mathbb{E}\left[\left(\exp\left\{X_T^{\varepsilon,H,H}\right\}
-e^x\right)^{+}\right]=\int_{x}^{\infty}(e^y-e^x)d\left[-P_{\varepsilon}^{0,H,T}(y)\right]. 
\label{E:BS}
\end{align}

Our next goal is to estimate the distribution function $P_{\varepsilon}^{0,H,T}(y)$. It follows from (\ref{E:vr}),
Chebyshev's exponential inequality, and the Cauchy-Schwartz inequality that for every $y> 0$,
\begin{align}
&P_{\varepsilon}^{0,H,T}(y)\le e^{-2y}
\mathbb{E}\left[\exp\left\{-\int_0^T\sigma(\varepsilon^{H}\widehat{B}_s)^2ds+2
\int_0^T\sigma(\varepsilon^{H}\widehat{B}_s)dZ_s\right\}\right] 
\nonumber \\
&\le e^{-2y}\mathbb{E}\left[\exp\left\{2\int_0^T\sigma(\varepsilon^{H}\widehat{B}_s)dZ_s\right\}\right] \nonumber \\
&=e^{-2y}\mathbb{E}\left[\exp\left\{-4\int_0^T\sigma(\varepsilon^{H}\widehat{B}_s)^2ds
+2\int_0^T\sigma(\varepsilon^{H}\widehat{B}_s)dZ_s+4\int_0^1\sigma(\varepsilon^{H}\widehat{B}_s)^2ds\right\}\right] 
\nonumber \\
&\le e^{-2y}\left(\mathbb{E}\left[\exp\left\{-8\int_0^1\sigma(\varepsilon^{H}\widehat{B}_s)^2ds
+4\int_0^T\sigma(\varepsilon^{H}\widehat{B}_s)dZ_s\right\}\right]\right)^{\frac{1}{2}} \nonumber \\
&\quad\times\left(\mathbb{E}\left[\exp\left\{8\int_0^T\sigma(\varepsilon^{H}\widehat{B}_s)^2ds\right\}\right]\right)^{\frac{1}{2}}.
\label{E:mart}
\end{align}
Now, using the linear growth condition for $\sigma$ and the fact that the stochastic exponential in (\ref{E:mart}) is a martingale
(see Lemma 13 in \cite{G}),
we obtain
\begin{align*}
&P_{\varepsilon}^{0,H,T}(y)\le e^{-2y}
\left(\mathbb{E}\left[\exp\left\{8\int_0^T\sigma(\varepsilon^{H}\widehat{B}_s)^2ds\right\}\right]\right)^{\frac{1}{2}} \\
&\le e^{-2y}e^{4c_1}\left(\mathbb{E}\left[\exp\left\{8c_2\varepsilon^{2H}\int_0^T\widehat{B}_s^2ds\right\}\right]\right)^{\frac{1}{2}}.
\end{align*}
It follows from (\ref{E:sti}) that there exists 
$\varepsilon_0> 0$ independent of $y$ and such that
\begin{equation}
\sup_{0<\varepsilon<\varepsilon_0}P_{\varepsilon}^{0,H,T}(y)\le l e^{-2y},
\label{E:matris}
\end{equation}
for some constant $l> 0$ independent of $y$. It is not hard to see that (\ref{E:BS}), (\ref{E:matris}), and the integration by parts formula imply the following:
\begin{align}
&C^{H,H,T}(\varepsilon,x)=\int_{x}^{\infty}e^y
P_{\varepsilon}^{0,H,T}(y)dy.
\label{E:otkr}
\end{align}
Next, using (\ref{E:otkr}), (\ref{E:disco}), (\ref{E:matris}), and the Lebesgue dominated convergence theorem, we see that 
for all $x> 0$, the equality in (\ref{E:integ}) holds.

We will next turn our attention to the case where $\alpha+\beta=H$ and $\beta\neq H$. This case is exceptional. It exhibits a
special discontinuity when compared with the neighboring regimes.
\begin{theorem}\label{T:verysp}
Suppose $\alpha+\beta=H$ and
$\beta\neq H$. Suppose also that the conditions in Corollary \ref{C:corio} hold, and the function $\sigma$ satisfies the linear growth condition. Then the following formula holds:
\begin{align*}
&C^{\beta,H,T}(\varepsilon,x\varepsilon^{\alpha})=
\varepsilon^{\alpha}\int_x^{\infty}\bar{{\cal N}}\left(\frac{y}{\sqrt{T}\sigma(0)}\right)dy+o\left(\varepsilon^{\alpha}\right)
\end{align*}
as $\varepsilon\downarrow 0$.
\end{theorem}

\it Proof. \rm It was established in the proof of Theorem \ref{T:clr} that for $\alpha=0$, 
$$
-\frac{1}{2}\varepsilon^{\alpha}\int_0^T\sigma(\varepsilon^{H}\widehat{B}_s)^2ds+
\int_0^T\sigma(\varepsilon^{H}\widehat{B}_s)dZ_s\rightarrow-\frac{1}{2}T\sigma(0)^2+\sigma(0)Z_T
$$
in probability. Making slight modifications, we can prove that for $\alpha\in(0,H]$,
$$
-\frac{1}{2}\varepsilon^{\alpha}\int_0^T\sigma(\varepsilon^{H}\widehat{B}_s)^2ds+
\int_0^T\sigma(\varepsilon^{H}\widehat{B}_s)dZ_s\rightarrow\sigma(0)Z_T
$$
in probability. Next, using the fact that convergence in probability implies convergence in distribution, we see that
\begin{equation}
\lim_{\varepsilon\downarrow 0}\mathbb{P}\left(X_T^{\varepsilon,\beta,H}\ge x\varepsilon^{\alpha}\right)
=\bar{{\cal N}}\left(\frac{x}{\sqrt{T}\sigma(0)}\right).
\label{E:du}
\end{equation}

We have
\begin{align}
&C^{\beta,H,T}(\varepsilon,x\varepsilon^{\alpha})=\mathbb{E}\left[\left(\exp\left\{X_T^{\varepsilon,\beta,H}\right\}
-\exp\left\{x\varepsilon^{\alpha}\right\}\right)^{+}\right] \nonumber \\
&=\int_{x\varepsilon^{\alpha}}^{\infty}\left(e^y-\exp\left\{x\varepsilon^{\alpha}\right\}\right)d\left[-P_{\varepsilon}^{\beta,H,T}(y)\right].
\label{E:BSO}
\end{align}
 It is not hard to see, by reasoning as in the proof of (\ref{E:matris}) that there exists $\varepsilon_1> 0$ such that
\begin{equation}
\sup_{0<\varepsilon<\varepsilon_1}P_{\varepsilon}^{\beta,H,T}(y)\le s e^{-2y},
\label{E:matrix}
\end{equation}
for some constant $s> 0$ and all $y> 0$. The estimate in (\ref{E:matrix}) allows us to integrate by parts in (\ref{E:BSO}).
This gives
\begin{align}
&C^{\beta,H,T}(\varepsilon,x\varepsilon^{\alpha})=\int_{x\varepsilon^{\alpha}}^{\infty}P_{\varepsilon}^{\beta,H,T}(y)e^ydy
=\varepsilon^{\alpha}\int_x^{\infty}\mathbb{P}\left(X_T^{\varepsilon,\beta,H}\ge u\varepsilon^{\alpha}\right)
\exp\left\{u\varepsilon^{\alpha}\right\}du.
\label{E:cur}
\end{align}
Next, using (\ref{E:matrix}) again, we can show that the dominated convergence theorem applies
to the integral in (\ref{E:cur}). Finally, taking into account (\ref{E:du}), (\ref{E:matrix}) and (\ref{E:cur}), we establish 
the asymptotic formula in Theorem \ref{T:verysp}.
\begin{remark}\label{R:rrr}
To the best of our knowledge, most of the methods, allowing to derive upper call price estimates in stochastic volatility models from small-time and small-noise large deviation principles for the log-price, rely either on the finiteness of nontrivial exponential moments of the integrated variance, or on the finiteness of the moments of the asset price for small values of the large deviation parameter. Some of the ideas used in such proofs go back to \cite{FJ}, Corollary 2.1, and \cite{Ph}, Subsection 5.1, where the Heston model was studied. It is worth mentioning that the exponential moment of order $p$ of the log-price in a Heston model is equal to the exponential moment of order $\frac{p(p-1)}{2}$ of the integrated variance in the Heston model with a different drift in the volatility equation (see formula (5.4) in \cite{Ph} and the formula preceding it). The finiteness of nontrivial exponential moments of the integrated variance plays an important role in some of the proofs of upper call price estimates in Gaussian stochastic volatility models. This can be seen, for instance, in the proof of Corollary 4.13 in \cite{FZ}, that of Corollary 31 in \cite{G}, and the proof of Theorem \ref{T:terl} in the present paper. However, Theorem \ref{T:confit} shows that this approach does not work if the volatility function grows at infinity faster than linearly. Moreover, in uncorrelated Volterra type models, the asset price process is a martingale, while all of its moments of order greater than one explode (see Theorem \ref{T:confitur}). An interesting method was used in \cite{FGPi}. It does not work for uncorrelated models, if the volatility function grows faster than linearly, but works well for correlated models under Assumption (A2) formulated in the next section (see the discussion after 
Corollary \ref{C:fina}).
\end{remark}
\section{The linear growth condition revisited}\label{S:revisit}
We have already advertized that the inequality
in (\ref{E:stihi}) may fail if the volatility function $\sigma$ grows at infinity a little faster than linearly (see Remark \ref{R:ooo}). In the present section, we
prove an assertion (Theorem \ref{T:confit}), which makes the previous statement more precise. This assertion states that under the faster than linear growth condition in Definition \ref{D:faster} below, all the positive order exponential moments of the quadratic variation of the driftless log-price are infinite. In addition, for uncorrelated Volterra type Gaussian stochastic volatility models, we establish that under the same growth condition, all the moments of order greater than one of the asset price are infinite (see part (i) of Theorem \ref{T:confitur}). Similar results for correlated models are obtained in part (ii) of Theorem \ref{T:confitur}.

\begin{definition}\label{D:faster}
Let $x_0> 0$. A positive function $f$ defined on $[x_0,\infty)$ grows faster than linearly if there exist $x_1> x_0$ and  
a positive function $g$ defined on $[x_1,\infty)$, for which the following conditions hold: $g(x)\rightarrow\infty$ as $x\rightarrow\infty$
and $f(x)\ge xg(x)$ for all $x> x_1$. 
\end{definition}

In the rest of this section, we will use certain results from the theory of slowly varying functions 
(see the monograph \cite{BGT} by Bingham, Goldie, and Teugels).
\begin{definition}\label{D:svf}
Let $l$ be a positive continuous function defined on some neighborhood of infinity and such that 
$$
\lim_{x\rightarrow\infty}\frac{l(\lambda x)}{l(x)}=1
$$
for all $\lambda> 0$. Any function $l$ satisfying the previous condition is called slowly varying.
\end{definition}

Definition \ref{D:svf} in a more general case of measurable functions can be found in Subsection 1.2.1 of \cite{BGT}. Slowly varying functions
were originally introduced and studied by Karamata in \cite{K}. The class of slowly varying functions is denoted by $R_0$.

Smoothly varying functions with index 0 play an important role in the theory of slow variation.
\begin{definition}\label{D:smvf}
A positive function $f$ defined on some neighborhood of infinity is called smoothly varying with index 0 
if the function $h(x)=\log f(e^x)$ is infinitely differentiable in a neighborhood of infinity and such that 
$h^{(n)}(x)\rightarrow 0$ 
as $x\rightarrow\infty$, for all integers $n\ge 1$.
\end{definition}
The class of all functions of smooth variation with index 0 is denoted by $SR_0$. It is known that 
$SR_0\subset R_0$. Moreover, the function
$f$ belongs to the class $SR_0$ if and only if
\begin{equation}
\lim_{x\rightarrow\infty}\frac{x^nf^{(n)}(x)}{f(x)}=0
\label{E:smva}
\end{equation}
for all $n\ge 1$. Definition \ref{D:smvf} in a more general case and the properties of functions of 
smooth variation formulated above can be found in Subsection 1.8.1
of \cite{BGT}. 

For positive continuous functions $f$ and $g$ defined on a neighborhood of infinity, the standard symbol $f\sim g$ will be used
when $\frac{f(x)}{g(x)}\rightarrow 1$ as $x\rightarrow\infty$. 
An important result in the theory of slow variation is the Smooth Variation Theorem (see Theorem 1.8.2 in \cite{BGT}). 
We will need only a special case of this theorem. A complete formulation can be found in \cite{BGT}.
\begin{theorem}\label{T:SVT}
Let $f\in R_0$. Then there exist $g\in SR_0$ and $h\in SR_0$ with $g\sim h$ and $g\le f\le h$ in a neighborhood of infinity.
\end{theorem}

The following statement follows from Theorem \ref{T:SVT}.
\begin{corollary}\label{C:SVTC}
Let $f\in R_0$ be such that $f(x)\rightarrow\infty$ as $x\rightarrow\infty$. Then there exist a function 
$g\in SR_0$ and a constant $x_0> 0$ such that $g$ is defined on $(x_0,\infty)$,
$g(x)\rightarrow\infty$ as $x\rightarrow\infty$, and $f(x)\ge g(x)$ for all $x> x_0$.
\end{corollary}

Let $f\in R_0$, and let the functions $g$ and $h$ be
such as in Theorem \ref{T:SVT}. To prove Corollary \ref{C:SVTC}, 
we only need to establish that $g(x)\rightarrow\infty$ as $x\rightarrow\infty$. It is easy to see that $f\sim g$, 
and the previous statement follows. 
\begin{remark}\label{R:BK}
It was established by Bojanic and Karamata (see \cite{BK}, see also Problem 11 on p. 124 in \cite{BGT}) that 
if $0< g(x)\rightarrow\infty$ as $x\rightarrow\infty$, then there exists $l\in R_0$ such that $0<l(x)\rightarrow\infty$
as $x\rightarrow\infty$ and $g(x)\ge l(x)$ for all $x> x_2$. It follows from the previous assertion that the function $g$
in Definition \ref{D:faster} can be replaced by a slowly varying function $l$ such that $0<l(x)\rightarrow\infty$
as $x\rightarrow\infty$. Moreover, the function $g$ can be replaced by a smoothly varying function $h$ such that 
$0<h(x)\rightarrow\infty$ as $x\rightarrow\infty$ (see Corollary \ref{C:SVTC}). We will use the previous remarks in the sequel.
\end{remark}

Our next goal is to prove an assertion confirming what was said in Remark \ref{R:ooo}.
\begin{theorem}\label{T:confit}
Suppose the volatility function $\sigma$ in the model described in (\ref{E:DD}) satisfies a faster than linear growth condition 
introduced in Definition \ref{D:faster}.
Let $\widehat{B}$ be a nondegenerate centered Gaussian process. 
Then, for all $t\in(0,T]$ and $\gamma> 0$, the following equality holds:
\begin{equation}
\mathbb{E}\left[\exp\left\{\gamma\int_0^t\sigma\left(\widehat{B}_s\right)^2ds\right\}\right]=\infty.
\label{E:lau}
\end{equation}
The equality in (\ref{E:lau}) also holds with the function $x\mapsto\sigma(-x)$ instead of the function $\sigma$.
\end{theorem}
\it Proof. \rm  Recall that we may assume that $\sigma(x)\ge xl(x)$ in a neighborhood of infinity, where $l$ is a slowly varying function such that in Remark \ref{R:BK}. It follows from the discussion above that there exist a number $x_2> x_1$ and a function $h\in SR_0$ 
defined on $(x_2,\infty)$ and such that $h(x)\rightarrow\infty$ as $x\rightarrow\infty$, 
and $l(x)^2\ge h(x)$ for all $x> x_2$. Define a function
$\hat{\sigma}(x)$ on $(x_2,\infty)$ by $\hat{\sigma}(x)=x^2h(x)$. Then $\sigma(x)^2\ge\hat{\sigma}(x)$ for all $x>x_2$.

Since $h$ is a strictly positive function, we have
$$
\frac{\hat{\sigma}^{\prime}(x)}{h(x)}=x\left(2+\frac{xh^{\prime}(x)}{h(x)}\right)
$$
and
$$
\frac{\hat{\sigma}^{\prime\prime}(x)}{h(x)}=2+2\frac{xh^{\prime}(x)}{h(x)}+\frac{x^2h^{\prime\prime}(x)}
{h(x)}.
$$
for all $x> x_2$. Next, using the condition $h\in SR_0$, the previous formulas, and (\ref{E:smva}), we see that there exists
$x_3> x_2$ such that the function $\hat{\sigma}$ is strictly increasing and convex in $[x_3,\infty)$.
Define a function $\tilde{\sigma}$ on $\mathbb{R}$ 
by the following: $\tilde{\sigma}(x)=\hat{\sigma}(x)$ if $x\ge x_3$, while $\tilde{\sigma}(x)=\hat{\sigma}(x_3)$ if $-\infty< x< x_3$.
It is not hard to see that the function $\tilde{\sigma}$ is convex in $\mathbb{R}$.

For every $x\in\mathbb{R}$, we have 
$
\sigma(x)^2\ge\tilde{\sigma}(x)-\hat{\sigma}(x_3).
$
Therefore, to establish (\ref{E:lau}), it suffices to prove that
$$
K=:\mathbb{E}\left[\exp\left\{\gamma\int_0^t\tilde{\sigma}\left(\widehat{B}_s\right)ds\right\}\right]=\infty.
$$

Since the function $\tilde{\sigma}$ is convex, Jensen's inequality implies that
\begin{equation}
\int_0^t\tilde{\sigma}\left(\widehat{B}_s\right)ds\ge t\tilde{\sigma}\left(\frac{1}{t}\int_0^t\widehat{B}_sds\right).
\label{E:J}
\end{equation}
It follows that for all $y> 0$, 
$$
K\ge e^{\gamma ty}\mathbb{P}\left(\tilde{\sigma}\left(\frac{1}{t}\int_0^t\widehat{B}_sds\right)> y\right).
$$
The function $\tilde{\sigma}$ is strictly increasing in $(x_3,\infty)$. Therefore, for $y> y_1$,
$$
K\ge e^{\gamma ty}\mathbb{P}\left(\int_0^t\widehat{B}_sds> t[\tilde{\sigma}]^{-1}(y)\right),
$$
and hence for $u> u_1$,
\begin{equation}
K\ge\exp\{\gamma t\tilde{\sigma}(u)\}\mathbb{P}\left(\int_0^t\widehat{B}_sds> tu\right)
=\exp\{\gamma tu^2h(u)\}\mathbb{P}\left(\int_0^t\widehat{B}_sds> tu\right).
\label{E:sit}
\end{equation}
The Riemann integral $\int_0^t\widehat{B}_sds$ of a nondegenerate continuous centered Gaussian process $\widehat{B}$ 
is a Gaussian random variable with mean zero and variance $v=\int_0^t\int_0^tC(u,s)duds$, where $C$ is the covariance 
function of the process $\widehat{B}$. Hence
$$
\mathbb{P}\left(\int_0^t\widehat{B}_sds> tu\right)=\frac{1}{\sqrt{2\pi}}\int_{\frac{tu}{\sqrt{v}}}^{\infty}
\exp\left\{-\frac{w^2}{2}\right\}dw
$$
with $v> 0$. Next, using the inequality
$$
\int_z^{\infty}e^{-\frac{u^2}{2}}du\ge e^{-\frac{z^2}{2}}\frac{2}{z+\sqrt{z^2+4}},
$$
which can be derived from 7.1.13 in \cite{AS}, we obtain
\begin{equation}
\mathbb{P}\left(\int_0^t\widehat{B}_sds> tu\right)\ge\frac{\sqrt{2}}{\sqrt{\pi}}\exp\left\{-\frac{t^2u^2}{2v}\right\}
\frac{\sqrt{v}}{tu+\sqrt{t^2u^2+4v}}.
\label{E:vi}
\end{equation}
Finally, by taking into account (\ref{E:sit}), (\ref{E:vi}) and the fact that $h(u)\uparrow\infty$ as $u\rightarrow\infty$,
we see that $K=\infty$. To prove the last statement in the formulation of Theorem \ref{T:confit} we apply the formula in (\ref{E:lau}) to the process $-\widehat{B}$.

This completes the proof of Theorem \ref{T:confit}.

We will next show that if the volatility function grows at infinity faster than the third power, then an assertion analogous to that in Theorem \ref{T:confit} can be established for the absolute value of the driftless log-price. 
\begin{theorem}\label{T:finalll}
Suppose the volatility function $\sigma$ appearing in (\ref{E:DD}) is such that
there exist a number $x_0> 0$ and a function $g$ for which the following conditions hold:
$0< g(x)\rightarrow\infty$ as $x\rightarrow\infty$ and
\begin{equation}
\sigma(x)\ge x^3g(x),\quad x> x_0.
\label{E:suppp}
\end{equation}
Let $\widehat{B}$ be a nondegenerate continuous Gaussian process adapted to the
filtration $\{\mathcal{F}_t\}_{0\le t\le T}$. 
Then, for all $t\in(0,T]$ and $\gamma> 0$,
\begin{equation}
\mathbb{E}\left[\exp\left\{\gamma\left|\int_0^t\sigma\left(\widehat{B}_s\right)dZ_s\right|\right\}\right]=\infty.
\label{E:lauta}
\end{equation}
The equality in (\ref{E:lauta}) also holds with the function $x\mapsto\sigma(-x)$ instead of the function $\sigma$.
\end{theorem}
\begin{remark}\label{R:ji}
Note that there is a gap between the linear growth condition in Definition \ref{D:faster} and the cubic growth condition in (\ref{E:suppp}). We do not know whether Theorem \ref{T:finalll}
holds true if condition (\ref{E:suppp}) in it is replaced by the condition in Definition \ref{D:faster} .
\end{remark}

\it Proof of Theorem \ref{T:finalll}. \rm It is clear from Remark \ref{R:BK} that we can replace the function $g$ in (\ref{E:suppp}) 
by a function $l\in R_0$ such that $l(x)\rightarrow\infty$ as $x\rightarrow\infty$. In addition, with no loss of generality, 
we may assume that 
\begin{equation}
\mathbb{E}\left[\int_0^T\sigma\left(\widehat{B}_s\right)^2ds\right]<\infty.
\label{E:maart}
\end{equation}
(expand the exponential in (\ref{E:lauta}) and use It$\rm\hat{o}$'s isometry). Set
\begin{equation}
M_t=\int_0^t\sigma\left(\widehat{B}_s\right)dZ_s,\quad 0\le t\le T. 
\label{E:lpo}
\end{equation}
Then the process $M$ is a martingale. Its quadratic variation is given by 
$$
[M]_t=\int_0^t\sigma\left(\widehat{B}_s\right)^2ds,\quad 0\le t\le T.
$$ 
Denote by $M^{*}$ the maximal function associated with the process $M$, that is
$$
M^{*}(t)=\sup_{0\le s\le t}|M(s)|,\quad 0\le t\le T.
$$
Then, using the Burkholder-Davis-Gundy inequality, more precisely, the lower estimate in it, with the constants such as in 
\cite{Ren}, Theorem 2, we obtain the following:
$$
\frac{c_1}{\sqrt{n}}||[M]^{\frac{1}{2}}(t)||_n\le||M^{*}(t)||_n,
$$
for all integers $n\ge 1$ and an absolute constant $c_1> 0$. In addition, using Doob's martingale inequality, we get
$$
||M^{*}(t)||_n\le\frac{n}{n-1}||M(t)||_n,\quad n\ge 2.
$$
Therefore, for all integers $n\ge 2$,
$$
\frac{c_2}{\sqrt{n}}||[M]^{\frac{1}{2}}(t)||_n\le||M(t)||_n,
$$
with an absolute constants $c_2> 0$.

Denote by $L_{\gamma,t}$ the expectation in (\ref{E:lauta}). Then the previous estimates give
\begin{align}
L_{\gamma,t}&\ge\mathbb{E}\left[\sum_{n=0}^{\infty}\frac{c_2^n\gamma^n}{n!n^{\frac{n}{2}}}
\left(\int_0^t\sigma\left(\widehat{B}_s\right)^2ds\right)
^{\frac{n}{2}}\right]+\gamma\mathbb{E}[|M(t)|]-c_2\gamma\mathbb{E}[[M]^{\frac{1}{2}}(t)] \nonumber \\
&\ge\mathbb{E}\left[\sum_{n=0}^{\infty}\frac{c_2^n\gamma^n}{n!n^{\frac{n}{2}}}
\left(\int_0^t\sigma\left(\widehat{B}_s\right)^2ds\right)
^{\frac{n}{2}}\right]-c_2\gamma\mathbb{E}\left[\left(\int_0^T\sigma(\widehat{B}_2)^2ds\right)^{\frac{1}{2}}\right]
\nonumber \\
&\ge\mathbb{E}\left[\sum_{n=0}^{\infty}\frac{c_2^n\gamma^n}{n!n^{\frac{n}{2}}}
\left(\int_0^t\sigma\left(\widehat{B}_s\right)^2ds\right)
^{\frac{n}{2}}\right]-C,
\label{E:ito}
\end{align}
where $C> 0$ does depend on $t$. It follows from (\ref{E:ito}) that in order to prove the equality $L_{\gamma,t}=\infty$,
it suffices to prove that
\begin{equation}
K_t:=\mathbb{E}\left[\sum_{n=0}^{\infty}\frac{c_2^n\gamma^n}{n!n^{\frac{n}{2}}}
\left(\int_0^t\sigma\left(\widehat{B}_s\right)^2ds\right)
^{\frac{n}{2}}\right]=\infty.
\label{E:nova}
\end{equation}

It follows from Stirling's formula that there exists 
$c_3> 0$ such that for every $n\ge 1$, $n^n\le c_3^nn!$. Therefore, (\ref{E:nova}) implies that for some $c_4> 0$,
$$
K_t\ge\mathbb{E}\left[\sum_{n=0}^{\infty}\frac{c_4^n\gamma^n}{(n!)^{\frac{3}{2}}}
\left(\int_0^t\sigma\left(\widehat{B}_s\right)^2ds\right)
^{\frac{n}{2}}\right].
$$
Let us set
\begin{equation}
\widetilde{K}_t:=\mathbb{E}\left[\left(\sum_{n=0}^{\infty}\frac{c_5^{\frac{2n}{3}}\gamma^{\frac{2n}{3}}}{n!}
\left(\int_0^t\sigma\left(\widehat{B}_s\right)^2ds\right)
^{\frac{n}{3}}\right)^{\frac{3}{2}}\right],
\label{E:eqw}
\end{equation}
where $c_5=\frac{c_4}{2}$. Next, applying H\"{o}lder's inequality with $p=\frac{3}{2}$ and $q=3$ to the sum in
(\ref{E:eqw}), we obtain
$$
\widetilde{K}_t\le\tau\mathbb{E}\left[\sum_{n=0}^{\infty}\frac{c_4^n\gamma^n}{(n!)^{\frac{3}{2}}}
\left(\int_0^t\sigma\left(\widehat{B}_s\right)^2ds\right)
^{\frac{n}{2}}\right]\le\tau K_t,
$$
where $\tau=\left\{\sum_{n=0}^{\infty}2^{-3n}\right\}^{\frac{1}{2}}$. Therefore, 
\begin{equation}
K_t\ge\frac{1}{\tau}
\mathbb{E}\left[\exp\left\{\frac{3}{2}c_5^{\frac{2}{3}}\gamma^{\frac{2}{3}}\left(\int_0^t\sigma\left(\widehat{B}_s\right)^2ds\right)
^{\frac{1}{3}}\right\}\right],
\label{E:poka}
\end{equation}
for all $t\in(0,T]$.
Note that we have not yet used the faster than cubic growth condition for the function $\sigma$. 
It follows from (\ref{E:poka}) that in order to finish the proof of Theorem \ref{T:finalll}, it suffices to show that 
$\widetilde{L}_{\gamma,t}=\infty$ for all $\gamma> 0$ and $t\in(0,T]$, where $\widetilde{L}_{\gamma,t}$ is the expectation on the right-hand side of (\ref{E:poka}) 
 
The rest of the proof of Theorem \ref{T:finalll} follows that of Theorem \ref{T:confit}. We first choose a function $h\in SR_0$ 
such that $h(x)\rightarrow\infty$ as $x\rightarrow\infty$, 
and moreover $l(x)^2\ge h(x)$ for all $x> x_2$. The function $x\mapsto x^6h(x)$ is strictly increasing and convex on $[x_3,\infty)$. 
Set $\sigma_0^2(x)=x^6h(x)\mathbb{1}{\{x> x_3\}}+x^6_3h(x_3)\mathbb{1}{\{x\le x_3\}}$, $x\in\mathbb{R}$. 
Then it is clear that the function $\sigma_0^2$ is convex on 
$\mathbb{R}$ and $\sigma(x)^2\ge\sigma_0(x)^2-x^6_3h(x_3)$ for all $x\in\mathbb{R}$. Next, using Jensen's inequality
as in the proof of (\ref{E:J}), we can estimate $\widetilde{L}_{\gamma,t}$ from below by an expression similar 
to the last expression in (\ref{E:sit}) with the function $h^{\frac{1}{3}}$ instead of the function $h$. Finally, taking into account Theorem \ref{T:confit}, we establish the equality $\widetilde{L}_{\gamma,t}=\infty$.

This completes the proof of Theorem \ref{T:finalll}.

\begin{corollary}\label{C:fina}
Suppose the conditions in Theorem \ref{T:finalll} hold. Then, for every $t\in(0,T]$, at least one of the next two conditions hold:
\begin{equation}
\mathbb{E}\left[\exp\left\{\gamma\int_0^t\sigma\left(\widehat{B}_s\right)dZ_s\right\}\right]=\infty\quad\mbox{for all}\quad\gamma> 0,
\label{E:filauta}
\end{equation}
or
\begin{equation}
\mathbb{E}\left[\exp\left\{-\gamma\int_0^t\sigma\left(\widehat{B}_s\right)dZ_s\right\}\right]=\infty\quad\mbox{for all}\quad\gamma> 0.
\label{E:filautar}
\end{equation}
\end{corollary}

\it Proof. \rm 
Using (\ref{E:lauta}) and the inequality 
$
e^{|u|}\le e^u+e^{-u},\,\,u\in\mathbb{R},
$
we see that for every $t\in(0,T]$ and $\gamma> 0$ either
\begin{equation}
\mathbb{E}\left[\exp\left\{\gamma\int_0^t\sigma\left(\widehat{B}_s\right)dZ_s\right\}\right]=\infty,
\label{E:filautarz}
\end{equation}
or
\begin{equation}
\mathbb{E}\left[\exp\left\{-\gamma\int_0^t\sigma\left(\widehat{B}_s\right)dZ_s\right\}\right]=\infty.
\label{E:filautaran}
\end{equation}

Fix $t> 0$. If there is no $\gamma$ for which (\ref{E:filautarz}) holds, 
then (\ref{E:filautaran}) should hold for all $\gamma> 0$ by the previous, which shows that (\ref{E:filautar}) is valid for such 
a number $t$. Otherwise set 
$$
a_t=\inf\{\gamma> 0:\mbox{(\ref{E:filautarz}) is true}\}.
$$ 
It is easy to see using H\"{o}lder's inequality that 
(\ref{E:filautarz}) is valid for all $\gamma> a_t$. Suppose $a_t=0$. Then (\ref{E:filauta}) holds. On the other hand, if $a_t> 0$, then
(\ref{E:filautaran}) is true for all $\gamma\in(0,a_t)$, and hence it is also true for all $\gamma> 0$. 

The proof of Corollary \ref{C:fina} is thus completed.

In \cite{BFGHS}, the authors consider Volterra type Gaussian stochastic volatility models, in which the asset price process $S$ satisfies the following conditions: \\
\\
(iiia) $S$ is a martingale; \\
\\
(iiib) For every $1<\gamma<\infty$ there exists $t> 0$ such that
$\mathbb{E}\left[S_t^{\gamma}\right]<\infty$ \\
\\(see Assumption 2.4 in Section 2 of \cite{BFGHS}). It was shown in \cite{BFGHS} that if conditions 
(iiia) and (iiib) hold, then upper large deviation style estimates for the call price follow from the corresponding large deviation principle.

An interesting improvement of the previous statement was obtained in \cite{FGPi}, where the same implication was obtained under weaker 
restrictions (see Assumption (A2) in \cite{FGPi}). In terms of the time parameter $t$, Assumption (A2) is as follows: There exists 
$\gamma> 1$ such that $\limsup_{t\rightarrow 0}\mathbb{E}[S_t^{\gamma}]<\infty$. It is not hard to see that if condition (iiia) holds, then 
Assumption (A2) follows from the following assumption: (i)\,There exist $\gamma> 1$ and $t> 0$ such that 
$\mathbb{E}\left[S_t^{\gamma}\right]<\infty$. 

In the rest of the present section, we consider Volterra type Gaussian stochastic volatility models. It will be shown next that for uncorrelated Volterra type models, Assumption (A2) does not hold if the volatility function $\sigma$ grows faster than linearly. More precisely, all the moments of order greater than one or less than zero of the asset price explode (see Theorem \ref{T:confitur}). 
Moreover, we prove in Theorem 
\ref{T:confitur} that for correlated models ($\rho\neq 0$), the moments explode for $\gamma\in(-\infty,0)\cup
(\frac{1}{1-\rho^2},\infty)$. In the case, where $\gamma=\frac{1}{1-\rho^2}$ in a correlated model, the moment explosion results obtained in the present section are only partial (see Theorem \ref{T:finaluli}).
Note that in the assertions established in the rest of the present section, we only assume that the Volterra type process $\widehat{B}$ satisfies the conditions in Definition \ref{D:Volt1}. Let us also recall that with no loss of generality we are assuming that $s_0=1$. 

The asset price process $S$ in an uncorrelated Gaussian stochastic 
volatility model with $s_0=1$ is given by
$$
S_t=\exp\left\{-\frac{1}{2}\int_0^t\sigma(\widehat{B}_s)^2ds+\int_0^t\sigma(\widehat{B}_s)dW_s\right\},\quad 0\le t\le T,
$$
Since Brownian
motions $W$ and $B$ are independent, it is rather standard to prove, by conditioning on the path of the process B, that the process $S$ is a martingale. 
Therefore, (iiia) holds true for the uncorrelated model. Moreover, for all $t\in[0,T]$, we have
\begin{equation}
\mathbb{E}\left[S_t\right]=1,\quad 0\le t\le T.
\label{E:rr}
\end{equation}
For negatively correlated models ($\rho< 0$), it was established in the paper \cite{Gassiat} of Gassiat that the asset price process $S$ is a martingale under the following additional condition: \\
\\
\it Condition (G): \rm For every $a\in\mathbb{R}$, the function $\sigma$ is bounded on $(-\infty,a]$. 
\\
\\
It follows that the equality in (\ref{E:rr}) holds for any negatively correlated Volterra type model, in which the volatility function 
$\sigma$ satisfies condition (G). On the other hand, for every $\eta\in\mathbb{R}$, the stochastic exponential
$$
t\mapsto\exp\left\{-\frac{\eta^2}{2}\int_0^t\sigma(\widehat{B}_s)^2ds+\eta\int_0^t\sigma(\widehat{B}_s)dB_s\right\}
$$
is a strictly positive local martingale. Hence it is a supermartingale, and therefore
\begin{equation}
\mathbb{E}\left[\exp\left\{-\frac{\eta^2}{2}\int_0^t\sigma(\widehat{B}_s)^2ds+\eta\int_0^t\sigma(\widehat{B}_s)dB_s\right\}\right]\le 1,
\label{E:ed}
\end{equation}
for every $t\in(0,T]$. Note that condition (G) is not needed to establish the validity of (\ref{E:ed}).

The next assertion is one of the main results of the paper. It concerns moment explosions in Gaussian stochastic volatility models
(see the discussion in the introduction, where we compare Theorem \ref{T:confitur} with the results obtained in \cite{Gassiat}).
\begin{theorem}\label{T:confitur}
(i)\,Suppose the volatility function $\sigma$ in an uncorrelated Volterra type Gaussian stochastic volatility model satisfies the faster than linear growth condition formulated in Definition \ref{D:faster}. Then the following equality holds:
\begin{equation}
\mathbb{E}\left[S_t^{\gamma}\right]=\infty\quad\mbox{for all $\gamma\in(-\infty,0)\cup(1,\infty)$ 
and $t\in(0,T]$}.
\label{E:hehe}
\end{equation}
(ii)\,Suppose $\rho\neq 0$, and suppose also that the faster than linear growth condition holds in a Volterra type Gaussian stochastic volatility model. Then, the following equality is valid:
\begin{equation}
\mathbb{E}\left[S_t^{\gamma}\right]=\infty\quad\mbox{for all $\gamma\in(-\infty,0)\cup(\frac{1}{1-\rho^2},\infty)$ 
and $t\in(0,T]$}.
\label{E:he}
\end{equation}
\end{theorem}

\it Proof. \rm It is clear that if $\rho=0$, then
\begin{align*}
&\mathbb{E}\left[S_t^{\gamma}\right]=\mathbb{E}\left[\exp\left\{\frac{\gamma^2-\gamma}{2}
\int_0^t\sigma(\widehat{B}_s)^2ds\right\}\exp\left\{-\frac{\gamma^2}{2}\int_0^t\sigma(\widehat{B}_s)^2ds
+\gamma\int_0^t\sigma(\widehat{B}_s)dW_s\right\}\right]. 
\end{align*}
Using the independence of $W$ and $B$ and conditioning on the path of the process B, we obtain
\begin{align*}
&\mathbb{E}\left[S_t^{\gamma}\right]=\mathbb{E}\left[\exp\left\{\frac{\gamma^2-\gamma}{2}
\int_0^t\sigma(\widehat{B}_s)^2ds\right\}{\cal E}_{t,\gamma}\right],
\end{align*}
where
\begin{equation}
{\cal E}_{t,\gamma}=\mathbb{E}\left[\exp\{-\frac{\gamma^2}{2}\int_0^t\sigma(\widehat{B}_s)^2ds
+\gamma\int_0^t\sigma(\widehat{B}_s)dW_s\}|B_s,0\le s\le t\right].
\label{E:st}
\end{equation}
It is not hard to prove that for all $t\in(0,T]$ and $\gamma$ such as in part (i) of the theorem, 
we have ${\cal E}_t=1$ a.s. Here we use the fact that the
simple stochastic exponentials appearing in (\ref{E:st}) are martingales.
It follows that
\begin{equation}
\mathbb{E}\left[S_t^{\gamma}\right]=\mathbb{E}\left[\exp\left\{\frac{\gamma^2-\gamma}{2}
\int_0^t\sigma(\widehat{B}_s)^2ds\right\}\right].
\label{E:22}
\end{equation}

Now suppose $\rho\neq 0$. Then the following equalities are true:
\begin{align*}
&\mathbb{E}\left[(S_t)^{\gamma}\right]=\mathbb{E}\left[\exp\left\{-\frac{\gamma}{2}\int_0^t\sigma\left(\widehat{B}_s\right)^2ds
+\gamma\int_0^t\sigma\left(\widehat{B}_s\right)dZ_s\right\}\right] \\
&=\mathbb{E}[\exp\{\frac{\gamma^2\bar{\rho}^2-\gamma}{2}\int_0^t\sigma\left(\widehat{B}_s\right)^2ds
+\gamma\rho\int_0^t\sigma\left(\widehat{B}_s\right)dB_s\} \\
&\quad\exp\{-\frac{\gamma^2\bar{\rho}^2}{2}\int_0^t\sigma\left(\widehat{B}_s\right)^2ds+\gamma\bar{\rho}
\int_0^t\sigma\left(\widehat{B}_s\right)dW_s\}].
\end{align*}
By conditioning on the path of the process $B$ and reasoning as in the proof of (\ref{E:22}), we obtain the following generalization of the formula in (\ref{E:22}):
\begin{equation}
\mathbb{E}\left[(S_t)^{\gamma}\right]=\mathbb{E}\left[\exp\left\{\frac{\gamma^2\bar{\rho}^2-\gamma}{2}
\int_0^t\sigma(\widehat{B}_s)^2ds+\gamma\rho\int_0^t\sigma\left(\widehat{B}_s\right)dB_s\right\}\right].
\label{E:alien}
\end{equation}

Let us first prove the equality in (\ref{E:hehe}). By taking into account the restriction $\gamma^2-\gamma> 0$ imposed in part (i) of Theorem \ref{T:confitur}, we see that (\ref{E:hehe}) follows from Theorem \ref{T:confit}. 

We will next prove (\ref{E:he}). Suppose the conditions in part (ii) of Theorem \ref{T:confitur} hold. Let $\eta\in\mathbb{R}$, and
fix $p> 1$ and $q> 1$ such that $\frac{1}{p}+\frac{1}{q}=1$. Then (\ref{E:alien}), (\ref{E:ed}), and 
H\"{o}lder's inequality imply the following estimate:
\begin{align}
&\left\{\mathbb{E}\left[S_t^{\gamma}\right]\right\}^{\frac{1}{p}}
\ge\left\{\mathbb{E}\left[S_t^{\gamma}\right]\right\}^{\frac{1}{p}}
\left\{\mathbb{E}\left[\exp\left\{-\frac{\eta^2}{2}\int_0^t\sigma(\widehat{B}_s)^2ds+\eta\int_0^t\sigma(\widehat{B}_s)dB_s\right\}\right]
\right\}^{\frac{1}{q}} \nonumber \\
&\ge\mathbb{E}\left[\exp\left\{\left(\frac{\gamma^2\bar{\rho}^2-\gamma}{2p}-\frac{\eta^2}{2q}\right)\int_0^t\sigma(\widehat{B}_s)^2ds
+\left(\frac{\gamma\rho}{p}+\frac{\eta}{q}\right)\int_0^t\sigma(\widehat{B}_s)dB_s\right\}\right].
\label{E:ooi}
\end{align}

Let us choose $\eta=-\frac{\gamma\rho}{p-1}$. Then $\frac{\gamma\rho}{p}+\frac{\eta}{q}=0$. 
Next, using (\ref{E:ooi}), we obtain
\begin{equation}
\left\{\mathbb{E}\left[S_t^{\gamma}\right]\right\}^{\frac{1}{p}}
\ge\mathbb{E}\left[\exp\left\{\left(\frac{\gamma^2\bar{\rho}^2-\gamma-\eta^2(p-1)}{2p}\right)\int_0^t\sigma(\widehat{B}_s)^2ds\right\}\right].
\label{E:ooio}
\end{equation}
It remains to choose $p> 1$ so that $l:=\gamma^2\bar{\rho}^2-\gamma-\eta^2(p-1)> 0$. Then (\ref{E:ooio}) and Theorem \ref{T:confit}
will show that the equality in \ref{E:he} holds.

We have $l=\gamma^2\bar{\rho}^2-\gamma-\frac{\gamma^2\rho^2}{p-1}$. If $\gamma>\frac{1}{1-\rho^2}$, then
$\gamma^2\bar{\rho}^2-\gamma> 0$, and the condition $l> 0$ is equivalent to the inequality
$p>1+\frac{\gamma\rho^2}{\gamma\bar{\rho}^2-1}$. Such a number $p> 1$ can be easily chosen. On the other hand, if $\gamma< 0$, then
the condition $l> 0$ is equivalent to the following inequality: $p> 1+\frac{\gamma^2\rho^2}{\gamma^2(1-\rho^2)+|\gamma|}$. As before, the previous condition allows us to easily choose $p$. It follows that (\ref{E:he}) holds.

The proof of Theorem \ref{T:confitur} is thus completed.
\begin{remark}\label{R:Gas}
We are indebted to Paul Gassiat for a suggestion to use the semimartingale property in (\ref{E:ed}) in the proof of part (ii) of Theorem
\ref{T:confitur}. Our original proof used condition (G) and the martingale property in (\ref{E:rr}).
\end{remark}

The next theorem complements Theorem \ref{T:confitur}. However, the conclusion in this theorem is weaker than that 
in Theorem \ref{T:confitur}.
\begin{theorem}\label{T:finaluli}
Suppose the volatility function $\sigma$ in a Volterra type Gaussian stochastic volatility model satisfies 
the faster than cubic growth condition formulated in (\ref{E:suppp}). Let $\rho\neq 0$, and  
let $\widetilde{{\cal M}}$ be the Volterra type model, in which the volatility function and the volatility process are the same as in the given model, while the correlation parameter is $-\rho$ instead of $\rho$. Denote by $\widetilde{S}$ the asset price process in the
model $\widetilde{{\cal M}}$.
Then, for every $t\in(0,T]$, at least one of the 
following two conditions holds: 
$$
\mathbb{E}\left[(S_t)^{\frac{1}{1-\rho^2}}\right]=\infty,\quad\mbox{or}\quad\mathbb{E}\left[(\widetilde{S}_t)
^{\frac{1}{1-\rho^2}}\right]=\infty.
$$
\end{theorem}

\it Proof. \rm It follows from (\ref{E:alien}) that if $\rho\neq 0$, then
$$
\mathbb{E}\left[(S_t)^{\frac{1}{1-\rho^2}}\right]=\mathbb{E}\left[\exp\left\{\frac{\rho}{1-\rho^2}
\int_0^t\sigma\left(\widehat{B}_s\right)dB_s\right\}\right].
$$
Similarly, we have
$$
\mathbb{E}\left[(\widetilde{S}_t)^{\frac{1}{1-\rho^2}}\right]=
\mathbb{E}\left[\exp\left\{-\frac{\rho}{1-\rho^2}\int_0^t\sigma\left(\widehat{B}_s\right)dB_s\right\}\right].
$$
Now, it is clear that Theorem \ref{T:finaluli} follows from Corollary \ref{C:fina}.

\begin{remark}\label{R:jourdain}
The condition $\gamma\ge\frac{1}{1-\rho^2}$ in the context of moment explosions appears in the paper \cite{J} of Jourdain, where the Scott model is studied. In the Scott model, the volatility function is $\sigma(x)=e^x$, while the volatility process is the 
Ornstein-Uhlenbeck process. To adapt Jourdain's results to our setting, we consider driftless Ornstein-Uhlenbeck processes. Jourdain proved
that for the Scott model with the driftless Ornstein-Uhlenbeck process as the volatility process, equality (\ref{E:he}) holds 
if $\rho=0$. Moreover, he established that if $\rho< 0$, then for given $t> 0$ and $\gamma> 1$, $\mathbb{E}[S_t^{\gamma}]
<\infty$ if and only if $\gamma<\frac{1}{1-\rho^2}$ (see Proposition 6 in \cite{J}). For $\rho> 0$, Jourdain proved that $\mathbb{E}[S_t^{\gamma}]=\infty$ if $\gamma\ge\frac{1}{1-\rho^2}$, and mentioned that it does not seem easy to analyze whether 
$\mathbb{E}[S_t^{\gamma}]<\infty$ if $\gamma<\frac{1}{1-\rho^2}$ (see Remark 7 in \cite{J}).
\end{remark}

\section{Asymptotic behavior of the implied volatility in mixed regimes}\label{S:varren}
In this section, we describe small-noise asymptotic behavior of the implied volatility in the mixed regimes 
considered in the previous section.

The implied volatility can be determined from the equality
\begin{align}
C^{\beta,H,T}\left(\varepsilon,x\varepsilon^{\alpha}\right)
&=C_{BS}(\varepsilon,x\varepsilon^{\alpha};\sigma=\widehat{\sigma}^{\beta,H,T}(\varepsilon,x\varepsilon^{\alpha})) \nonumber \\
&=C_{-}(x\varepsilon^{\alpha},\sqrt{\varepsilon}\widehat{\sigma}^{\beta,H,T}(\varepsilon,x\varepsilon^{\alpha})).
\label{E:inne}
\end{align}

In the cases, where $0\le\alpha+\beta< H$, Theorem \ref{T:terl} implies that
\begin{equation}
L(\varepsilon)=:\log\frac{1}{C_{-}(x\varepsilon^{\alpha},\sqrt{\varepsilon}\widehat{\sigma}^{\beta,H,T}(\varepsilon,x\varepsilon^{\alpha}))}
=J_T(x)\varepsilon^{2\alpha+2\beta-2H}
+o\left(\varepsilon^{2\alpha+2\beta-2H}\right)
\label{E:firus}
\end{equation}
as $\varepsilon\downarrow 0$. In the previous formula, the symbol $J_T$ stands for the rate function $I_T$ defined in (\ref{E:vunz}),
in the case where $\alpha+\beta=0$ (here we assume that the assumptions in part (i) of Theorem \ref{T:terl} hold), while
$J_T(x)=\frac{x^2}{2T\sigma(0)^2}$, in the case where $0<\alpha+\beta< H$, and the assumptions in part (ii) of Theorem \ref{T:terl} hold. 
In (\ref{E:firus}), the parametrized dimensionless implied volatility is given by $\nu(\varepsilon)
=\sqrt{\varepsilon}\widehat{\sigma}^{\beta,H,T}(\varepsilon,x\varepsilon^{\alpha})$.
Moreover, we have 
$$
\frac{k(\varepsilon)}{L(\varepsilon)}=O\left(\varepsilon^{2H-\alpha-2\beta}\right)
$$
as $\varepsilon\rightarrow 0$. Therefore, $\frac{k(\varepsilon)}{L(\varepsilon)}\rightarrow 0$ as $\varepsilon\rightarrow 0$. This means that the formula in Remark 7.3
in \cite{GL} can be applied to characterize the asymptotic behavior of the dimensionless implied volatility 
$\varepsilon\mapsto\nu(\varepsilon)$
in the mixed regime. In our case, the formula in \cite{GL}, Remark 7.3, gives the following:
$$
\left|\frac{k(\varepsilon)^2}{2L(\varepsilon)}-\varepsilon\widehat{\sigma}^{\beta,H}(\varepsilon,x\varepsilon^{\alpha})^2\right|=
o\left(\frac{k(\varepsilon)^2}{L(\varepsilon)}\right)
$$
as $\varepsilon\downarrow 0$. It follows that
\begin{equation}
\left|\frac{k(\varepsilon)}{\sqrt{2L(\varepsilon)}}-\sqrt{\varepsilon}\widehat{\sigma}^{\beta,H}(\varepsilon,x\varepsilon^{\alpha})\right|=
o\left(\frac{k(\varepsilon)}{\sqrt{L(\varepsilon)}}\right)
\label{E:RL}
\end{equation}
as $\varepsilon\downarrow 0$. Next, taking into account (\ref{E:firus}), we obtain the following assertion.
\begin{theorem}\label{T:terul}
(i)\,\,Suppose the conditions in Theorem \ref{T:1} hold. Suppose also that $\alpha+\beta=0$, and the linear growth
condition holds for the function $\sigma$. Then
$$
\widehat{\sigma}^{\beta,H,T}(\varepsilon,x\varepsilon^{\alpha})=\frac{x}{\sqrt{2I_T(x)}}\varepsilon^{H-\beta-\frac{1}{2}}
+o\left(\varepsilon^{H-\beta-\frac{1}{2}}\right)
$$
as $\varepsilon\downarrow 0$. \\
(ii)\,\,Suppose the conditions in Corollary \ref{C:corio} hold. Suppose also that $0<\alpha+\beta< H$, and the linear growth
condition holds for the function $\sigma$. Then
$$
\widehat{\sigma}^{\beta,H,T}(\varepsilon,x\varepsilon^{\alpha})=\sqrt{T}\sigma(0)\varepsilon^{H-\beta-\frac{1}{2}}
+o\left(\varepsilon^{H-\beta-\frac{1}{2}}\right)
$$
as $\varepsilon\downarrow 0$. 
\end{theorem}

Let $\alpha=0$ and $\beta=H$. Then, for the Black-Scholes model with $\sigma=\widehat{\sigma}^{H,H,T}(\varepsilon,x)$, the equality in 
(\ref{E:otkr}) takes the following form:
\begin{align}
&C_{BS}(\varepsilon,x;\widehat{\sigma}^{H,H,T}(\varepsilon,x)) \nonumber \\
&=\int_{x}^{\infty}e^y\bar{{\cal N}}\left(\frac{y}
{\sqrt{\varepsilon}\widehat{\sigma}^{H,H,T}(\varepsilon,x))}
+\frac{1}{2}\sqrt{\varepsilon}\widehat{\sigma}^{H,H,T}(\varepsilon,x)\right)dy.
\label{E:cro}
\end{align}
Using (\ref{E:inne}), (\ref{E:integ}), and (\ref{E:cro}), we obtain
\begin{align}
&\int_{x}^{\infty}e^y\bar{{\cal N}}\left(\frac{y}{\sqrt{T}\sigma(0)}+\frac{\sqrt{T}\sigma(0)}{2}\right)dy \nonumber \\
&=\lim_{\varepsilon\downarrow 0}\int_{x}^{\infty}e^y\bar{{\cal N}}\left(\frac{y}
{\sqrt{\varepsilon}\widehat{\sigma}^{H,H,T}(\varepsilon,x))}
+\frac{1}{2}\sqrt{\varepsilon}\widehat{\sigma}^{H,H,T}(\varepsilon,x)\right)dy,
\label{E:rach1}
\end{align}
for all $x> 0$.

Let $\varepsilon_j$, $j\ge 1$, be a positive sequence such that $\varepsilon_j\rightarrow 0$ as $j\rightarrow\infty$, and the limit 
$$
\tau=\lim_{j\rightarrow\infty}\sqrt{\varepsilon_j}\widehat{\sigma}^{H,H,T}(\varepsilon_j,x)
$$ 
exists (finite or infinite). Applying Fatou's lemma to the expression on the right-hand side of (\ref{E:rach1}) and taking into account the fact that the call price function $C_{-}$ is strictly increasing in $\nu$ (see Remark \ref{R:oio}), we see that $\tau\le\sigma(0)$. Therefore, for $j\ge j_0$, 
$$
\sqrt{\varepsilon_j}\widehat{\sigma}^{H,H,T}(\varepsilon_j,k=x)\le C,
$$ 
where $C> 0$ is a constant, and hence we have
$$
\sup_{j\ge j_0}\left[e^y\bar{{\cal N}}\left(\frac{y}
{\sqrt{\varepsilon_j}\widehat{\sigma}^{H,H,T}(\varepsilon_j,x))}
+\frac{1}{2}\sqrt{\varepsilon_j}\widehat{\sigma}^{H,H,T}(\varepsilon_j,x)\right)\right]\le e^y\bar{{\cal N}}\left(\frac{y}{C}\right).
$$
The previous estimate allows us to
apply the dominated convergence theorem in formula (\ref{E:rach1}) (along the sequence $\varepsilon_j$). This gives
$C_{-}(x,\sqrt{T}\sigma(0))=C_{-}(x,\tau)$, and hence $\tau=\sqrt{T}\sigma(0)$.
Now, it is clear that
$$
\lim_{\varepsilon\downarrow 0}\sqrt{\varepsilon}\widehat{\sigma}^{H,H,T}(\varepsilon,x)=\sqrt{T}\sigma(0).
$$
Therefore, the following statement holds.
\begin{theorem}\label{T:ivoh}
Suppose $\alpha=0$ and $\beta=H$. Then, under the assumptions in Corollary \ref{C:corio}
and the linear growth condition,
$$
\widehat{\sigma}^{H,H,T}(\varepsilon,x)=\sqrt{T}\sigma(0)\varepsilon^{-\frac{1}{2}}+o\left(\varepsilon^{-\frac{1}{2}}\right)
$$
as $\varepsilon\downarrow 0$. 
\end{theorem}

We will next turn our attention to the only remaining case of the implied volatility estimates in mixed regimes. 
\begin{theorem}\label{T:nakonets}
Suppose $\alpha+\beta=H$ and $\alpha\in(0,H]$. Then, under the assumptions in Corollary \ref{C:corio}
and the linear growth condition, the following asymptotic formula holds for the implied volatility:
\begin{equation}
\widehat{\sigma}^{\beta,H,T}(\varepsilon,x\varepsilon^{\alpha})=\frac{x\varepsilon^{\alpha-\frac{1}{2}}}
{\sqrt{2\alpha\log\frac{1}{\varepsilon}}}
+o\left(\frac{\varepsilon^{\alpha-\frac{1}{2}}}{\sqrt{\log\frac{1}{\varepsilon}}}\right)
\label{E:fora}
\end{equation}
as $\varepsilon\downarrow 0$.
\end{theorem}

\it Proof. \rm It follows from Theorem \ref{T:verysp} that
$$
L(\varepsilon)=\log\frac{1}{C^{\beta,H,T}(\varepsilon,x\varepsilon^{\alpha})}=\alpha\log\frac{1}{\varepsilon}
-\log\int_x^{\infty}\bar{{\cal N}}\left(\frac{y}{\sqrt{T}\sigma(0)}\right)dy+o(1)
$$
as $\varepsilon\downarrow 0$. We also have $k(\varepsilon)=x\varepsilon^{\alpha}$, and hence $\frac{k(\varepsilon)}
{L(\varepsilon)}\rightarrow 0$ as $\varepsilon\downarrow 0$. Next, applying the formula in Remark 7.3
in \cite{GL} (see (\ref{E:RL}) above), we derive (\ref{E:fora}).

The proof of Theorem \ref{T:nakonets} is thus completed.

\section*{Acknowledgements}\label{S:A}
I thank Jean-Dominique Deuschel, Peter Friz, Josselin Garnier, Paul Gassiat, Stefan Gerhold, Benjamin Jourdain, Barbara Pacciarotti, Paolo Pigato, and Knut S\o lna for their attention to the paper and valuable remarks.

\end{document}